\documentclass[12pt]{article}

\usepackage{amsmath}

\makeatother

\newcommand{\hs}{\hspace}
\newcommand{\ts}{\hspace*}
\newcommand{\vs}{\vspace}
\newcommand{\e}{\enskip}
\newcommand{\q}{\quad}

\newcommand{\dps}{\displaystyle}
\newcommand{\f}{\frac}

\newcommand{\Phat}{\hat{{\cal P}}}

\newcommand{\PhatB}{\hat{{\bf P}}}
\newcommand{\Qhat}{\hat{{\cal Q}}}

\newcommand{\Omhat}{\hat{\Omega}}

\newcommand{\hxi}{\hat{\xi}}
\newcommand{\hpi}{\hat{\pi}}

\newcommand{\Zp}{\hat{Z}^{(+)}}
\newcommand{\Zm}{\hat{Z}^{(-)}}
\newcommand{\Zpm}{\hat{Z}^{(\pm)}}
\newcommand{\Zmp}{\hat{Z}^{(\mp)}}

\newcommand{\xip}{\hat{\xi}^{(+)}}
\newcommand{\xim}{\hat{\xi}^{(-)}}
\newcommand{\pip}{\hat{\pi}^{(+)}}
\newcommand{\pim}{\hat{\pi}^{(-)}}

\newcommand{\xipp}[1]{\hat{\xi}^{\mbtn{#1}(+)}}
\newcommand{\ximm}[1]{\hat{\xi}^{\mbtn{#1}(-)}}
\newcommand{\pipp}[1]{\hat{\pi}^{\mbtn{#1}(+)}}
\newcommand{\pimm}[1]{\hat{\pi}^{\mbtn{#1}(-)}}

\newcommand{\xips}[1]{\hat{\xi}^{(+)#1}}
\newcommand{\xims}[1]{\hat{\xi}^{(-)#1}}
\newcommand{\pips}[1]{\hat{\pi}^{(+)#1}}
\newcommand{\pims}[1]{\hat{\pi}^{(-)#1}}

\newcommand{\Zmms}[2]{\hat{Z}^{\mbtn{#1}(-)}_{#2}}

\newcommand{\bM}{\bar{M}}
\newcommand{\mcK}{\mathcal{K}}
\newcommand{\mcC}{\mathcal{C}}
\newcommand{\mcA}{\mathcal{A}}

\newcommand{\mcS}{\mathcal{S}}
\newcommand{\mcH}{\mathcal{H}}

\newcommand{\mcG}{\mathcal{G}}
\newcommand{\mcM}{\mathcal{M}}

\newcommand{\mcU}{\mathcal{U}}

\newcommand{\mcV}{\mathcal{V}}

\newcommand{\mcB}{\mathcal{B}}

\newcommand{\vLd}{\varLambda}

\newcommand{\tM}{\tilde{M}}
\newcommand{\tX}{\tilde{X}}

\newcommand{\commu}[2]{[#1,#2]}
\newcommand{\commut}[2]{[#1,\e #2]}
\newcommand{\symmp}[2]{\ \{#1,\e #2\}}
 
\newcommand{\mbtn}[1]{\mbox{{\tiny #1}}}
\newcommand{\pscrp}{\mbox{{\scriptsize $\Phat$}}}

\newcommand{\ACCS}{{\it ACCS}}
\newcommand{\hop}{{\it hyper}-operator}

\makeatletter
 
\@addtoreset{equation}{section}
\makeatother

\begin{document}

\begin{center}
{\bf \Large Uncertainty Relations and Quantum Corrections in Noncommutative Quantum Mechanics on a Curved Space
}\vs{12pt}\\
M. Nakamura\footnote{\label{*}E-mail:mnakamur@hm.tokoha-u.ac.jp}
\vs{12pt}\\
{\it Research Institute, Hamamatsu Campus, Tokoha University, Miyakoda-cho 1230, 
Kita-ku, Hamamatsu-shi, Shizuoka 431-2102, Japan}
\end{center}

\begin{abstract}
Starting with the first-order singular Lagrangian describing the dynamical system with 2nd-class constraints, the noncommutative quantum mechanics on a curved space is investigated by the constraint star-product quantization formalism of the projection operator method. Imposing the additional constraints to eliminate the reduntant degrees of freedom, it is shown that the resultant noncommutative quantum system on the curved space is represented with two kinds of the constrained quantum systems, which are equivalent with each other. Then, it is shown that the resultant Hamiltonians contain the quantum corrections caused by the uncertainty relations among the constraint-operators  in addition to those due to the projections of operators, which are missed in the usual approaches with the Dirac-bracket quantization formalism. 
\end{abstract}

\section{Introduction}

\ts{12pt}The problem of the noncommutative extensions of the quantum systems constrained to a submanifold embedded in the higher-dimensional Euclidean space has been investigated widely investigated as one of the quantum theories on a curved space untill now\cite{S1,S2}. As the curved space, the submanifold $M^{N-1}$ specified by $G(x)=0$ ($G(x)\in {\it C}^{\infty}$) in an $N$-dimensional Euclidean space $R^N$ has been considered in many studies, where $x=(x^1,\cdots,x^i,\cdots,x^N)\in R^N$. Then, we have shown in the previous studies\cite{S2} that the projected constrained quantum systems contain the quantum corrections associated to the projections of operators through the constraint star-product quantization formalism of projection operator method(POM)\cite{S3,S4,S5,S6}. 
\\
\ts{12pt}As shown in our previous studies\cite{S3,S7,S8}, the POM satisfies the decomposition of unity formula for the associated canonically conjugate set (ACCS) of the constraint operators. From this formula, then, we will propose the {\it ACCS}-expansion formula in the POM.
\\
\ts{12pt}In this paper, we will construct {\it exactly} the noncommutative quantum system on a curved space in the general form. Then, it will be shown that the commutator-algebras and the Hamiltomians in the resultant constraint quantum systems contain the quantum corrections associated to the uncertainty relations among the constraint-operators in addition to those due to the projections of operators, which are missed in the usual approaches with the Dirac-bracket quantization formalism. 
\\
\ts{12pt}The present paper is qrganized as follows. In Sect.2, we propose the brief review of the constraint star-product quantization formalism of the POM and the {\it ACCS}-expansion formulas. In Sect.3, we set up the initial unconstraint quantum system, and the consistent-set of constraint-operators and the Lagrange multiplier operators are fixed. Imposing the additional constraints, in Sect.4, the resultant noncommutative quantum systems on the curved space are constructed in the {\it exact} form, and the quantum corrections in these resultant systems are investigated. In Sect.5, the discussion and the some concluding remarks are given. 

\section{{\it ACCS}-Expansion of Constraint System}

\ts{12pt}Following the previous works \cite{S2}, we here present the brief review of the constraint star-product quantization formalism and the {\it ACCS}-expansion formulas in quantum constraint systems.
\subsection{Star-product quantization}

\ts{12pt}Let $\mcS=(\mcC,\mcA(\mcC),H(\mcC),\mcK)$ be the initial unconstraint quantum system, where $\mcC=\{(q^i,p_i);i=1,\cdots ,N\}$ is a set of canonically conjugate operators (CCS), $\mcA(\mcC)$, the commutator algebra of $\mcC$ defined with  
$$
\mcA(\mcC):\commu{q^i}{p_j}=i\hbar\delta^i_j,\hs{36pt}\commu{q^i}{q^j}=\commu{p_i}{p_j}=0,
\eqno{(2.1)}
$$ 
and $H(\mcC)$ is the Hamiltonian of the initial unconstraint system, $\mcK=\{T_{\alpha}(\mcC)|\alpha=1,\cdots ,2M<2N\}$, the set of the constraint-operators $T_{\alpha}(\mcC)$ corresponding to the second-class constraints $T_{\alpha}\approx 0$. Starting with $\mcS$, our task is to construct the constraint quantum system  
$\mcS^*=(\mcC^*,\mcA(\mcC^*),H^*(\mcC^*))$, where $\mcC^*$ is the projected CCS satisfying
$$
T_{\alpha}(\mcC^*)=0\hs{48pt}(\alpha=1,\cdots,2M).
\eqno{(2.2)}
$$
For this purpose, we first construct the associated canonically conjugate set (\ACCS) from $\mcK$ and the projection operator $\Phat$, which is defined as the \hop, to eliminate $T_{\alpha}\ (\alpha=1,\cdots,2M)$. Then, $\mcC^*$ is defind by $\mcC^*=\Phat\mcC$.\\
\ts{12pt}Let $\{(\xi^a,\pi_ a)|a=1,\cdots,M\}$ be the ACCS, and their symplectic forms be
$$ 
Z_{\alpha}=\left\{\begin{array}{l}\xi^a\hs{24pt}(\alpha=a)\vs{6pt}\\
\pi_a\hs{24pt}(\alpha=a+M) \hs{36pt}(\alpha=1,\cdots,2M\q;\q a=1,\cdots,M),
\end{array}\right.
\eqno{(2.3)}
$$

which obey the commutator algebra
$$
\begin{array}{l}
\commut{\xi^a}{\pi_b}=i\hbar\delta^a_b,\hs{12pt}\commut{\xi^a}{\xi^b}=
\commut{\pi_a}{\pi_b}=0,\vs{6pt}\\
\commut{Z_{\alpha}}{Z_{\beta}}=i\hbar J^{\alpha\beta},
\end{array}
 \eqno{(2.4)}
$$ 
where $J^{\alpha\beta}$ is the $2N\times2N$ symplectic matrix.\\
\ts{12pt}We next define the symplectic {\it hyper}-operators $\Zpm_{\alpha}(\alpha=1,\dots,2M)$ as follows:\footnote{For any operators $A,B$, $\symmp{A}{B}=\dps{\f12}(AB+BA)$.} 
$$
\Zm_{\alpha}=\f1{i\hbar}\commu{Z_{\alpha}}{\q},\hs{36pt}\Zp_{\alpha}=\symmp{Z_{\alpha}}{\q}.
\eqno{(2.5)}
$$
From (2.4), $\Zpm$ obey the {\it hyper}-commutator algebra
$$
\begin{array}{l}
[\Zpm_{\alpha},\Zpm_{\beta}]=0,\vs{6pt}\\
\commu{\Zpm_{\alpha}}{\Zmp_{\beta}}=\commu{\Zmp_{\alpha}}{\Zpm_{\beta}}=J^{\alpha\beta}.
\end{array}
\eqno{(2.6)}
$$
Then, the projection operator $\Phat$ is defined by
$$
\Phat=\exp\left[(-1)^s\Zp_{\alpha}\f{\partial}{\partial\varphi_{\alpha}}\right]\exp[J^{\alpha\beta}\varphi_{\alpha}\Zm_{\beta}]|_{\varphi=0},
\eqno{(2.7)}
$$  
which satisfies the projection conditions 
$$
\Phat T_{\alpha}(\mcC)=T_{\alpha}(\mcC^*)=0\hs{48pt}(\alpha=1,\cdots,2M)
\eqno{(2.8)}
$$
and the following formulas for the decomposion of unity:
$$
\begin{array}{lcl}
\hat{{\bf I}}&=&\exp\left[-(-1)^s\Zp_{\alpha}\f{\partial}{\partial\varphi_{\alpha}}\right]\Phat\exp[J^{\alpha\beta}\varphi_{\alpha}\Zm_{\beta}]|_{\varphi=0}\vs{12pt}\\
&=&
\exp\left[(-1)^s\Zp_{\alpha}\f{\partial}{\partial\varphi_{\alpha}}\right]\Phat\exp[-J^{\alpha\beta}\varphi_{\alpha}\Zm_{\beta}]|_{\varphi=0},
\end{array}
\eqno{(2.9)}
$$
where $\hat{{\bf I}}=1$ is the unity \hop.\\
\ts{12pt}The {\it hyper}-operator $\Omhat_{\eta\zeta}$ in the constraint star-product quantization formalism is defined by
$$
\Omhat_{\eta\zeta}=J^{\alpha\beta}\Zm_{\alpha}(\eta)\Zm_{\beta}(\zeta)=\hxi^{a}(\eta)\hpi_a(\zeta)-\hpi_a(\eta)\hxi^{a}(\zeta)
\eqno{(2.10)}
$$
with the nonlocal representations for the operations of \hop s, which satisfies
$$
\Omhat^t_{\eta\zeta}=\Omhat_{\zeta\eta}=-\Omhat_{\eta\zeta},
\eqno{(2.11)}
$$
and two-kinds of star-product are defined as follows:
$$
X\star Y=\left.\exp(\f{\hbar}{2i}\Omhat_{\eta\zeta})X(\eta)Y(\zeta)\right|_{\eta=\zeta}
\eqno{(2.12)}
$$  
and
$$
X\pscrp\star   Y=\left.\left(\Phat(\eta)\Phat(\zeta)\exp(\f{\hbar}{2i}\Omhat^t_{\eta\zeta})X(\eta)Y(\zeta)\right)\right|_{\eta=\zeta}.
\eqno{(2.13)}
$$
\ts{12pt}Using the $\star$ and $\pscrp\star$-products, the commutator-formulas and the symmetrized product-ones under the operation of $\Phat$ are expressed as follows:
$$
\begin{array}{lcl}
\commu{\Phat X}{\Phat Y}&=&\Phat\commu{X}{Y}_{\star}=\Phat(X\star Y-Y\star X),\vs{12pt}\\
\symmp{\Phat X}{\Phat Y}&=&\Phat\symmp{X}{Y}_{\star}=\dps{\f12}\Phat(X\star Y+Y\star  X),
\end{array}
\eqno{(2.14a)}
$$  
and
$$
\begin{array}{lcl}
\Phat\commu{X}{Y}&=&\commu{X}{Y}_{\pscrp\star}=(X\pscrp\star Y-Y\pscrp\star X),\vs{12pt}\\
\Phat\symmp{X}{Y}&=&\symmp{X}{Y}_{\pscrp\star}=\dps{\f12}(X\pscrp\star Y+Y\pscrp\star X).
\end{array}
\eqno{(2.14b)}
$$
\subsection{{\it ACCS}-expansion of operators}

From the formula (2.9), any operator $O(\mcC)$ is represented in the following form\footnote{$
\xips{n}\pips{m}=\xip_{a_1}\cdots\xip_{a_n}\pip_{b_1}\cdots\pip_{b_m},\hs{24pt}\xims{m}\pims{n}=\xim_{b_m}\cdots\xim_{b_1}\pim_{a_n}\cdots\pim_{a_1}.
$}:
$$
\begin{array}{lcl}
O(\mcC)&=&\dps{\hat{{\bf I}}  O(\mcC)=\sum^{\infty}_{n=0}\f1{n!}J^{\alpha_1\beta_1}\cdots J^{\alpha_n\beta_n}\Zp_{\alpha_1}\cdots\Zp_{\alpha_n}\Phat\Zm_{\beta_n}\cdots\Zm_{\beta_1}O(\mcC)}\vs{12pt}\\

&=&\dps{\sum^{\infty}_{n=0,m=0}\f{(-1)^n}{n!m!}\xips{n}\pips{m}\Phat\xims{m}\pims{n}O(\mcC)}\vs{12pt}\\
&=&\Phat O(\mcC)+O'(Z,\mcC),
\end{array}
\eqno{(2.15a)}
$$ 
where 
$$
O'(Z,\mcC)=\sum^{\infty}_{n,m=0;n+m\neq 0}
\f{(-1)^n}{n!m!}\xips{n}\pips{m}\Phat\xims{m}\pims{n}O(\mcC)
\eqno{(2.15b)}
$$ 
\ts{12pt}In the decomposition of $O(\mcC)$, Eq.(2.15a), the projected part $\Phat O(\mcC)$ contains the quantum correction terms caused by the operator ordering, and the {\it ACCS}-expansion part $O'(Z,\mcC)$ products the other type of quantum corrections associated to the uncertainty relations for the {\it ACCS}
$$
\Delta\xi_a\Delta\pi_b\geq\f{\hbar}2\delta_{ab}.
\eqno{(2.16)}
$$
\ts{12pt}From the decomposition (2.15a), the initial Hilbert space $\mcH$ is defined as follows:
$$
\mcH=\mcH^*\oplus\mcH^c,
\eqno{(2.17)}
$$ 
where $\mcH^*$ is the subspace with the CCS $\mcC^*$, and $\mcH^c$, that with the ACCS $\{Z_{\alpha}|\alpha=1,\cdots,2M\}$. Then, the \hop\ $\PhatB$ is defined by
$$
\PhatB O(\mcC)=\f{<\Phi|O(\mcC)|\Phi>}{<\Phi|\Phi>}
\eqno{(2.18)}
$$
with $\Phi\in \mcH^c$, which satisfies the following formulas:
$$
\PhatB\PhatB=\PhatB,\hs{12pt}\Zm_{\alpha}\PhatB=0,\hs{12pt}\Phat\PhatB=\PhatB,\hs{12pt}\PhatB\Phat=\Phat.
\eqno{(2.19)}
$$
\ts{12pt}Using the \hop\ $\hat{{\bf I}}$ and $\PhatB$, $O(\mcC)$ is projected out into the constraint subspace $\mcH^*$ in the following form:
$$
\begin{array}{lcl}

O(\mcC)&\rightarrow&O^*(\mcC)=\PhatB\hat{{\bf I}}O(\mcC)\vs{12pt}\\

&=&\dps{\Phat O(\mcC)+\sum^{\infty}_{n,m=0;n+m\neq 0}
\f{(-1)^n}{n!m!}<\xips{n}\pips{m}1>_{\Phi}\Phat\xims{m}\pims{n}O(\mcC)}\vs{12pt}\\

&=&\Phat O+\Qhat O

\end{array}
\eqno{(2.20a)}
$$ 
with $<\xips{n}\pips{m}1>_{\Phi}=\PhatB\xips{n}\pips{m}1$, where 
$$
\begin{array}{lcl}
\Phat O&=&\Phat O(\mcC),\vs{12pt}\\
\Qhat O&=&\dps{\sum^{\infty}_{n,m=0;n+m\neq 0}
\f{(-1)^n}{n!m!}<\xips{n}\pips{m}1>_{\Phi}\Phat\xims{m}\pims{n}O(\mcC)},
\end{array}
\eqno{(2.20b)}
$$
and, $\Phi$ is one of the several relevant states to minimize the uncertainty relations among $Z$'s. As such a state, we shall take the ground state of the coherent states 
with respect to the $\ACCS$ $\{(\xi,\pi)\}$,
which is denoted with $\Phi^c$, and is defined by 
$$
<\xi|\Phi^c>=\left(\f1{\pi\hbar}\right)^{1/4}\exp(-\f1{2\hbar}\xi_a\xi_a)
\eqno{(2.21)}
$$
in the Schr\"{o}dinger representation. Using (2.21), the fundamental expectation values for $\ACCS$ with respect to $\Phi^c$ become as follows:
$$
\begin{array}{rcl}
<\Phi^c|\xip_a\cdot 1|\Phi^c>&=&<\Phi^c|\pip_a\cdot 1|\Phi^c>=0,\vs{12pt}\\
<\Phi^c|\xip_a\xip_b\cdot 1|\Phi^c>&=&<\Phi^c|\pip_a\pip_b\cdot 1|\Phi^c>=\dps{\f{\hbar}2\delta_{ab}},\vs{12pt}\\
<\Phi^c|\xip_a\pip_b\cdot 1|\Phi^c>&=&<\Phi^c|\pip_a\xip_b\cdot1|\Phi^c>=0.
\end{array}
\eqno{(2.22)}
$$
Then, $\Qhat O$ is given as
$$
\Qhat O=\sum^{\infty}_{n,m=0;n+m\neq 0}
\f1{n!m!}\left(\f{\hbar}4\right)^{n+m}\Phat\xims{2m}\pims{2n}O(\mcC),
\eqno{(2.23)}
$$
which contains the quantum effects associated to the uncertainty relations among the $\ACCS$.\\
\ts{12pt}Thus, $O^*(\mcC)$ is represented in the following way:
$$
\begin{array}{lcl}
O^*(\mcC)&=&\PhatB\hat{{\bf I}}O(\mcC)=\Phat O(\mcC)+\Qhat O(\mcC)\vs{12pt}\\

&=&\Phat O(\mcC)+\dps{\sum^{\infty}_{n,m=0;n+m\neq 0}
\f1{n!m!}\left(\f{\hbar}4\right)^{n+m}\Phat\xims{2m}\pims{2n}O(\mcC)}.
\end{array}
\eqno{(2.24)}
$$ 

\section{Initial Hamiltonian System $\mcS$}

\ts{12pt}Let $\Theta$ be the totally antisymmetric tensor defined by
$$
\Theta^{ij}=\theta\varepsilon^{ij}
\eqno{(3.1)}
$$ 
with the constant noncommutative-parameter $\theta$ and the completely antisymmetric tensor  $\varepsilon^{ij}$ ($\varepsilon^{ij}=1\hs{6pt}(i>j),\hs{6pt}\varepsilon^{ji}=-\varepsilon^{ij}\hs{6pt}(i,j=1,\cdots,N)$), we shall consider the dynamical system described by the first-order singular Lagrangian \cite{S2} 
$$
\begin{array}{rcl}
L&=&L(x,\dot{x},v,\dot{v},\lambda,\dot{\lambda},u,\dot{u})\vs{12pt}\\

&=&\dot{x}^iv_i-\lambda\dot{G}(x)-\f12\dot{u}_i\Theta^{ij}u_j-h_0(x,v),
\end{array}
\eqno{(3.2a)}
$$
where $\dot{G}(x)=\dot{x}^iG_i(x)$\footnote[8]{$G_i(x)=\partial^x_{i}G(x)$ with $\partial^x_{i}= \partial/\partial x^i$.}, and $h_0(x,v)$ corresponds to the Hamiltonian of free particles,
$$
h_0(x,v)=\f12v_iv_i.
\eqno{(3.2b)}
$$
\ts{12pt}Then, the initial unconstraint quantum system $\mcS=(\mcC,\mcA(\mcC),H(\mcC),\mcK)$ is constructed as follows: \vs{12pt}\\
{\bf \boldmath i) Initial canonically conjugate set $\mcC$}
$$ 
\ts{-72pt}\mcC=\{(x^i,p^x_i),(v_i,p_v^i),(\lambda,p_{\lambda}),(u_i,p_u^i)|i=1,\cdots,N\},
\eqno{(3.3)}
$$
which obeys the commutator algebra $\mcA(\mcC)$:
$$
\begin{array}{l}
\commut{x^i}{p^x_j}=i\hbar\delta^i_j,\hs{12pt}\commut{v_i}{p_v^j}=i\hbar\delta_i^j,\hs{12pt}\commut{u_i}{p_u^j}=i\hbar\delta_i^j,\vs{12pt}\\
\commut{\lambda}{p_{\lambda}}=i\hbar,\hs{12pt}
\mbox{(the others)}=0.
\end{array}
\eqno{(3.4)}
$$
{\bf \boldmath ii) Initial Hamiltonian $H$}
$$
H=\symmp{\mu^i_{(1)}}{\phi^{(1)}_i}+\symmp{\mu^i_{(2)}}{\phi^{(2)}_i}+\symmp{\mu_{(3)}}{\phi^{(3)}}+h_0(x,v),
\eqno{(3.5)}
$$
where $\phi^{(n)}$, $(n=1,\cdots, 3)$ are the constraint operators corresponding to the primary constraints together with $\phi^{(4)}_i$ $(i=1,\cdots,N)$ and $\mu^i_{(n)}$ $(n=1,\cdots,4)$ are the Lagrange multiplier operators.
\vs{6pt}\\
{\bf \boldmath iii) Consistent set of constraints and the Lagrange multiplier operators}
\vs{6pt}\\
\ts{12pt}Through the consistency conditions for the time evolusions of constraint operators, the consistent set of constraints, $\mcK$, is set up as follows:
$$
\mcK=\{\phi^{\mbtn{(1)}}_i,\phi^{\mbtn{(2)}}_i,\phi^{\mbtn{(3)}},\phi^{\mbtn{(4)}}_i,\psi^{\mbtn{(1)}}\}
\eqno{(3.6)}
$$
with 
$$
\begin{array}{lcl}
\phi^{\mbtn{(1)}}_i=v_i-p^x_i-\lambda G_i(x),&\hs{6pt}&
\phi^{\mbtn{(2)}}_i=p_v^i,\vs{12pt}\\
\phi^{\mbtn{(3)}}=p_{\lambda},&\hs{6pt}&
\phi^{\mbtn{(4)}}_i=p_u^i+\dps{\f12}\Theta^{ij}u_j,\vs{12pt}\\
\psi^{\mbtn{(1)}}=G_i(x)v_i,
\end{array}
\eqno{(3.7)}
$$
where $\phi^{\mbtn{(n)}}\ (n=1,\cdots,4)$ are the constraint operators corresponding to the primary constraints and $\psi^{\mbtn{(1)}}$, one corresponding to the secondary constraint.\\
\ts{12pt}Then, the Lagrange multiplier operators, $\mu^i_{(1)}$, $\mu^i_{(2)}$, $\mu_{(3)}$ and $\mu^i_{(4)}$ are obtained with
$$
\begin{array}{lcl}
\mu_{\mbtn{(1)}}^i=-v_i,&\hs{6pt}&\mu_{\mbtn{(2)}}^i=\mu^{\mbtn{(2)}}_{i;kl}(x)v_kv_l,\vs{12pt}\\
\mu_{\mbtn{(3)}}=\mu^{\mbtn{(3)}}_{kl}(x)v_kv_l,&\hs{6pt}&\mu_{\mbtn{(4)}}^i=0,
\end{array}
\eqno{(3.8)}
$$
where
$$
\begin{array}{l}
\mcG(x)=G_i(x)G_i(x),\vs{12pt}\\
\mu^{{\mbtn{(2)}}}_{i;kl}(x)=-\mcG^{-1}(x)G_i(x)G_{kl}(x),\vs{12pt}\\
\mu^{{\mbtn{(3)}}}_{kl}(x)=-\mcG^{-1}(x)G_{kl}(x),
\end{array}
\eqno{(3.9)}
$$
which satisfies
$$
\mu^{{\mbtn{(2)}}}_{i;kl}(x)=G_i(x)\mu^{{\mbtn{(3)}}}_{kl}(x).
\eqno{(3.10)}
$$
\ts{12pt}The consistent set $\mcK$ obeys the commutator algebra $\mcA(\mcK)$:
$$
\begin{array}{lcl}

\commut{\phi^{\mbtn{(1)}}_i}{\phi^{\mbtn{(2)}}_j}=i\hbar\delta_{ij},&\hs{6pt}&\commut{\phi^{\mbtn{(2)}}_i}{\psi^{\mbtn{(1)}}}=-i\hbar G_i(x),\vs{12pt}\\

\commut{\phi^{\mbtn{(1)}}_i}{\phi^{\mbtn{(3)}}}=-i\hbar G_i(x),&\hs{6pt}&\commut{\phi^{\mbtn{(4)}}_i}{\phi^{\mbtn{(4)}}_j}=i\hbar\Theta^{ij},\vs{12pt}\\

\commut{\phi^{\mbtn{(1)}}_i}{\psi^{\mbtn{(1)}}}=i\hbar G_{ij}(x)v_j,&\hs{6pt}&\mbox{(the others)}=0.

\end{array}
\eqno{(3.11)}
$$
\ts{12pt}Thus, we have constructed the initial unconstraint quamtum system $\mcS$.

\section{Sequential Projections for $\mcK$}

\ts{12pt}Starting with the initial quantum system $\mcS$, we shall construct the constraint quantum system $\mcS^*$ strictly satisfying $\mcK=0$ through  the {\it ACCS}-expansion formulation in the star-produt quantization formalism with POM. 

\subsection{Classification of $\mcK$ and Sequential projections }

\ts{12pt}From the structure of the commutator algebra (3.11), we shall  classify $\mcK$ into the following three subsets:

$$
\begin{array}{l}

\mcK=\mcK^{\mbtn{(A)}}\oplus\mcK^{\mbtn{(B)}}\oplus\mcK^{\mbtn{(C)}},\vs{12pt}\\

\mcK^{\mbtn{(A)}}=\{\phi^{\mbtn{(1)}}_i,\phi^{\mbtn{(2)}}_i|i=1,\cdots N\},\hs{12pt}

\mcK^{\mbtn{(B)}}=\{\phi^{\mbtn{(3)}},\psi^{\mbtn{(1)}}\},\hs{12pt}

\mcK^{\mbtn{(C)}}=\{\phi^{\mbtn{(4)}}_i|i=1,\cdots,N\}.

\end{array}
\eqno{(4.1)}
$$
\ts{12pt}Taking account of the commutator algebra (3.11), then, the sequential projections of $\mcS$ can be uniquely carried out through the following projection-diagram:
$$
\mcS\rightarrow\mcS^{\mbtn{(1)}}=\mcS(\mcK^{\mbtn{(A)}}=0)\rightarrow\mcS^{\mbtn{(2)}}=\mcS^{\mbtn{(1)}}(\mcK^{\mbtn{(B)}}=0)\rightarrow\mcS^{\mbtn{(3)}}=\mcS^{\mbtn{(2)}}(\mcK^{\mbtn{(B)}}=0)\rightarrow\mcS^*,
\eqno{(4.2a)}
$$
where
$$
\mcS^{\mbtn{(1)}}=(\mcC^{\mbtn{(1)}},H^{\mbtn{(1)}},\mcK^{\mbtn{(B)}}\oplus\mcK^{\mbtn{(C)}}),\hs{12pt}\mcS^{\mbtn{(2)}}=(\mcC^{\mbtn{(2)}},H^{\mbtn{(2)}},\mcK^{\mbtn{(C)}}),\hs{12pt}\mcS^{\mbtn{(3)}}=(\mcC^{\mbtn{(3)}},H^{\mbtn{(3)}},\mcK=0).
\eqno{(4.2b)}
$$

\subsection{Construction of $\mcS^{\mbtn{(1)}}$}

Using the POM and the {\it ACCS}-expansion formulation, we shall provide $\mcS^{\mbtn{(1)}}$ with the precise form.

\subsubsection{{\it ACCS} for $\mcK^{\mbtn{(A)}}$}

From the commutator algebra $\mcA(\mcK)$, (3.11), the {\it ACCS} for $\mcK^{\mbtn{(A)}}$ is defined as
$$
Z^{\mbtn{(1)}}_{\alpha}=\left\{
                               \begin{array}{ll}
                                               \xi^{\mbtn{(1)}}_i=\phi^{\mbtn{(1)}}_i=v_i-p^x_i-\lambda G_i(x),&(\alpha=i),\vs{6pt}\\
                                               \pi^{\mbtn{(1)}}_i=\phi^{\mbtn{(2)}}_i=p_v^i&(\alpha=i+N).
                               \end{array}
                         \right.
\eqno{(4.3)}
$$
Then, $Z^{\mbtn{(1)}(-)}_{\alpha}$ operates on $\mcC$ in the following way:
$$
\begin{array}{lclclcl}
\ximm{(1)}_kx_i=\delta_{ki},&\hs{6pt}&\ximm{(1)}_k\lambda=0,&\hs{6pt}&\pimm{(1)}_kx^i=0,&\hs{6pt} &\pimm{(1)}_k\lambda=0,\vs{6pt}\\

\ximm{(1)}_kp^x_i=-\lambda G_{ki}(x),& &\ximm{(1)}_kp_{\lambda}=-G_k(x),& &\pimm{(1)}_kp^x_i=0,& &\pimm{(1)}_kp_{\lambda}=0,\vs{6pt}\\

\ximm{(1)}_kv_i=0,& &\ximm{(1)}_ku_i=0,& &\pimm{(1)}_kv_i=-\delta_{ki},& &\pimm{(1)}_ku_i=0,\vs{6pt}\\

\ximm{(1)}_kp_v^i=\delta_{ki},& &\ximm{(1)}_kp_u^i=0& &\pimm{(1)}_kp_v^i=0,& &\pimm{(1)}_kp_u^i=0,\vs{6pt}\\
& & & & & &(k=1,\cdots,N).

\end{array}
\eqno{(4.4)}
$$
\subsubsection{Projection operator $\Phat^{\mbtn{(1)}}$ and the projected CCS $\mcC^{\mbtn{(1)}}$}

\ts{12pt}Let the projection operator for $\mcK^{\mbtn{(A)}}$ be $\Phat^{\mbtn{(1)}}=\Phat(Z^{\mbtn{(1)}(+)}_{\alpha},Z^{\mbtn{(1)}(-)}_{\alpha})$, which satisfies the projection conditions : 
$$
\begin{array}{l}
\Phat^{\mbtn{(1)}}\phi^{\mbtn{(1)}}_i=\Phat^{\mbtn{(1)}}\xi^{\mbtn{(1)}}_i=0,\vs{12pt}\\

\Phat^{\mbtn{(1)}}\phi^{\mbtn{(2)}}_i=\Phat^{\mbtn{(1)}}\pi^{\mbtn{(1)}}_i=0.
\end{array}
\eqno{(4.5)}
$$
Then, the projected CCS $\mcC^{\mbtn{(1)}}$ is defined by 
$$
\begin{array}{lcl}
\mcC^{\mbtn{(1)}}&=&\Phat^{\mbtn{(1)}}\mcC=\{(\Phat^{\mbtn{(1)}}x^i,\Phat^{\mbtn{(1)}}p^x_i),(\Phat^{\mbtn{(1)}}v_i,\Phat^{\mbtn{(1)}}p^v_i),(\Phat^{\mbtn{(1)}}\lambda,\Phat^{\mbtn{(1)}}p_{\lambda}),(\Phat^{\mbtn{(1)}}u_i,\Phat^{\mbtn{(1)}}p^u_i)\}\vs{12pt}\\

&=&\{x^i,p^x_i,v_i,\lambda,p_{\lambda},u_i,p_u^i\},\vs{12pt}\\

\mbox{with}& &v_i-p^x_i-\lambda G_i(x)=0,\hs{6pt}p_v^i=0,

\end{array}
\eqno{(4.6)}
$$
which satisfies the commutator-algebra $\mcA(\mcC^{\mbtn{(1)}})$:
$$
\begin{array}{lcl}

\commut{x^i}{p^x_j}=i\hbar\delta^i_j,& &\commut{x^i}{v_j}=i\hbar\delta^i_j,\vs{6pt}\\

\commut{v_i}{p^x_j}=i\hbar\lambda G_{ij}(x),& &\commut{v_i}{p_{\lambda}}=i\hbar G_i(x),\vs{6pt}\\

\commut{\lambda}{p_{\lambda}}=i\hbar,& &\commut{u_i}{p_u^j}=i\hbar\delta_i^j,\hs{36pt}\mbox{(the others)}=0. 

\end{array}
\eqno{(4.7)}
$$
\vs{6pt}\\
The remaining constraints $\mcK^{\mbtn{(B)}}$ and $\mcK^{\mbtn{(C)}}$ are projected on to $\mcC^{\mbtn{(1)}}$ as follows:
$$
\begin{array}{lclcl}

\Phat^{\mbtn{(1)}}\phi^{\mbtn{(3)}}&=&\Phat^{\mbtn{(1)}}p_{\lambda}=p_{\lambda}=\phi^{\mbtn{(3)}}&\in&\mcC^{\mbtn{(1)}},\vs{12pt}\\

\Phat^{\mbtn{(1)}}\psi^{\mbtn{(1)}}&=&\Phat^{\mbtn{(1)}}\symmp{G_i(x)}{v_i}=\symmp{\Phat^{\mbtn{(1)}}G_i(x)}{\Phat^{\mbtn{(1)}}v_i}& &\vs{6pt}\\
&=&\symmp{G_i(x)}{v_i}=\psi^{\mbtn{(1)}}&\in&\mcC^{\mbtn{(1)}},\vs{12pt}\\

\Phat^{\mbtn{(1)}}\phi^{\mbtn{(4)}}&=&\Phat^{\mbtn{(1)}}(p^i_u+\dps{\f12}\Theta^{ij}u_j)=p^i_u+\dps{\f12}\Theta^{ij}u_j=\psi{\mbtn{(4)}}&\in&\mcC^{\mbtn{(1)}}.

\end{array}
\eqno{(4.8)}
$$
Thus,
$$
\begin{array}{lclcl}

\mcK^{\mbtn{(B)}}(\in\mcC^{\mbtn{(1)}})&=&\{\phi^{\mbtn{(3)}},\psi^{\mbtn{(1)}}\}&\mbox{with}&\mcA(\mcK^{\mbtn{(B)}}):\commut{\psi^{\mbtn{(1)}}}{\phi^{\mbtn{(3)}}}=i\hbar\mcG(x),\vs{12pt}\\

\mcK^{\mbtn{(C)}}(\in\mcC^{\mbtn{(1)}})&=&\{\phi^{\mbtn{(4)}}\}&\mbox{with}&\mcA(\mcK^{\mbtn{(C)}}):\commut{\phi^{\mbtn{(4)}}_i}{\phi^{\mbtn{(4)}}_j}=i\hbar\Theta^{ij}.

\end{array}
\eqno{(4.9)}
$$

\subsubsection{Projected Hamiltonian $H^{\mbtn{(1)}}$}

\ts{12pt}From the formula (2.24), $H^{\mbtn{(1)}}$ is constructed as follows:
$$
H^{\mbtn{(1)}}=\PhatB^{\mbtn{(1)}}\hat{{\bf I}}^{\mbtn{(1)}}H=\Phat^{\mbtn{(1)}}H+\Qhat^{\mbtn{(1)}}H,
\eqno{(4.10)}
$$
where
$$
\begin{array}{l}

\Phat^{\mbtn{(1)}}H\vs{6pt}\\

\dps{=\f12v_iv_i+\f{\hbar^2}4\mcG^{-1}(x)G_{i}(x)G_{ij}(x)+\symmp{\symmp{\mu^{\mbtn{(3)}}_{kl}(x)}{v_kv_l}}{p_{\lambda}}+\f{\hbar^2}4\symmp{\mu^{\mbtn{(3)}}_{kl}(x)_{;kl}}{p_{\lambda}}}

\end{array}
\eqno{(4.11a)}
$$
and 
$$
\begin{array}{l}

\Qhat^{\mbtn{(1)}}H=\dps{\sum^{\infty}_{n+m\neq 0}\f1{n!m!}\left(\f{\hbar}4\right)^{n+m}\Phat^{\mbtn{(1)}}(\pimm{(1)}_k\pimm{(1)}_k)^m(\ximm{(1)}_l\ximm{(1)}_l)^n}\ H\vs{6pt}\\

=-\dps{\f{\hbar}4}N+\mcU_{\mbtn{I}}(x)+\symmp{\mcU^{kl}_{\mbtn{II}}(x)}{v_kv_l}

+\symmp{\mcU_{\mbtn{III}}(x)}{p_{\lambda}}+\symmp{\symmp{\mcU^{kl}_{\mbtn{IV}}(x)}{v_kv_l}}{p_{\lambda}}.

\end{array}
\eqno{(4.11b)}
$$
The explicit forms of $\mcU_{\mbtn{I}}(x),\mcU_{\mbtn{II}}^{kl}(x),\mcU_{\mbtn{III}}(x)$ and $\mcU_{\mbtn{IV}}^{kl}(x)$ in $\Qhat^{\mbtn{(1)}}H$ are presented in Appendix A.\\
\ts{12pt}Then, $H^{\mbtn{(1)}}$ is represented in the following form:
$$
\begin{array}{lcl}

H^{\mbtn{(1)}}&=&-\dps{\f{\hbar}4N+\f12v_iv_i}\vs{12pt}\\

&+&U_{\mbtn{I}}(x)+\symmp{U_{\mbtn{II}}^{kl}(x)}{v_kv_l}+\symmp{U_{\mbtn{III}}(x)}{p_{\lambda}}+\symmp{\symmp{U_{\mbtn{IV}}^{kl}(x)}{v_kv_l}}{p_\lambda},

\end{array}
\eqno{(4.12a)}
$$
where
$$
\begin{array}{l}

U_{\mbtn{I}}(x)=\mcU_{\mbtn{I}}(x)+\dps{\f{\hbar^2}4}\mcG^{-1}(x)G_{kl}(x)G_{kl}(x),\vs{6pt}\\

U_{\mbtn{II}}^{kl}(x)=\mcU_{\mbtn{II}}^{kl}(x),\vs{6pt}\\

U_{\mbtn{III}}(x)=\mcU_{\mbtn{III}}(x)+\dps{\f{\hbar^2}4}\mu^{\mbtn{(3)}}_{kl}(x)_{;kl},\vs{6pt}\\

U_{\mbtn{IV}}^{kl}(x)=\mcU_{\mbtn{IV}}^{kl}(x)+\mu^{\mbtn{(3)}}_{kl}(x).

\end{array}
\eqno{(4.12b)}
$$
\ts{12pt}Thus, we have constructed the projected quantum system $\mcS^{\mbtn{(1)}}$:
$$
\mcS^{\mbtn{(1)}}=(\mcC^{\mbtn{(1)}},H^{\mbtn{(1)}},\mcK^{\mbtn{(B)}}\oplus\mcK^{\mbtn{(C)}}).
\eqno{(4.13)}
$$

\subsection{Construction of $\mcS^{\mbtn{(2)}}$}

\ts{12pt}Following the projection diagram (4.2a), we shall construct $\mcS^{\mbtn{(2)}}$, where $\mcK^{\mbtn{(B)}}=0$.

\subsubsection{{\it ACCS} of $\mcK^{\mbtn{(B)}}$}

From the commutator algebra $\mcA(\mcK^{\mbtn{(B)}})$ in (4.9),  
the {\it ACCS} is given by
$$
Z^{\mbtn{(2)}}_{\alpha}=\left\{
                               \begin{array}{ll}
                                               \xi^{\mbtn{(2)}}=\symmp{\mcG^{-1}(x)}{\psi^{\mbtn{(1)}}}=\symmp{\mcG^{-1}(x)G_i(x)}{v_i},&(\alpha=1),\vs{6pt}\\
                                               \pi^{\mbtn{(2)}}=\phi^{\mbtn{(3)}}=p_{\lambda}&(\alpha=i+N).
                               \end{array}
                         \right.
\eqno{(4.14)}
$$
Then, $Z^{\mbtn{(2)}(-)}_{\alpha}$ operates on $\mcC^{\mbtn{(1)}}$ as follows:
$$
\begin{array}{l}
\ximm{(2)}x_i=\nu_i(x),\hs{6pt}\ximm{(2)}\lambda=0,\hs{6pt}\ximm{(2)}p_{\lambda}=1,\hs{6pt}\ximm{(2)}u_i=0,\hs{6pt}\ximm{(2)}p_u^i=0,\vs{6pt}\\

\ximm{(2)}p^x_i=2\xipp{(2)}\mu^{{\mbtn{(2)}}}_{k;ki}(x)-\symmp{\mu^{{\mbtn{(3)}}}_{ik}(x)}{v_k}-\lambda\mu^{{\mbtn{(2)}}}_{k;ki}(x),\vs{6pt}\\

\ximm{(2)}v_i=2\xipp{(2)}\mu^{{\mbtn{(2)}}}_{k;ki}(x)-\symmp{\mu^{{\mbtn{(3)}}}_{ik}(x)}{v_k},\vs{6pt}\\

\pimm{(2)}v^i=-G_i(x),\hs{6pt}\pimm{(2)}\lambda=-1,\hs{12pt}\mbox{(the others)}=0,

\end{array}
\eqno{(4.15)}
$$ 
where
$$
\nu_i(x)=-\mcG^{-1}(x)G_i(x).
\eqno{(4.16)}
$$

\subsubsection{Projection operator $\Phat^{\mbtn{(2)}}$ and the projected CCS  $\mcC^{\mbtn{(2)}}$}

\ts{12pt}Let the projection operator for $\mcK^{\mbtn{(B)}}$ be $\Phat^{\mbtn{(2)}}=\Phat(Z^{\mbtn{(2)}(+)}_{\alpha},Z^{\mbtn{(2)}(-)}_{\alpha})$, which satisfies the projection conditions for $\mcK^{\mbtn{(B)}}$: 
$$
\begin{array}{l}
\Phat^{\mbtn{(2)}}\phi^{\mbtn{(3)}}=\Phat^{\mbtn{(2)}}p_{\lambda}=0,\vs{12pt}\\

\Phat^{\mbtn{(2)}}\psi^{\mbtn{(1)}}=\Phat^{\mbtn{(2)}}\xipp{(2)}\mcG(x)=0.
\end{array}
\eqno{(4.17)}
$$
Then, $\Phat^{\mbtn{(2)}}\mcC^{\mbtn{(1)}}$ becomes  
$$
\begin{array}{lcl}
\Phat^{\mbtn{(2)}}\mcC^{\mbtn{(1)}}&=&\{x^i,p^x_i,v_i,\lambda,u_i,p_u^i|i=1\cdots,N\},\vs{6pt}\\

\mbox{with}& &v_i-p^x_i-\symmp{\lambda}{G_i(x)}=0,\vs{6pt}\\

& &p_v^i=0,\hs{6pt}p_{\lambda}=0,\hs{6pt}\symmp{G_i(x)}{v_i}=0,
\end{array}
\eqno{(4.18)}
$$
which obeys the commutator-algebra $\mcA(\Phat^{\mbtn{(2)}}\mcC^{\mbtn{(1)}})$:
$$
\begin{array}{l}

\commut{x^i}{p^x_j}=i\hbar\delta^i_j,\vs{6pt}\\

\commut{u_i}{p_u^j}=i\hbar\delta_i^j,\vs{6pt}\\
 \commut{v_i}{v_j}=i\hbar\symmp{\mu^{{\mbtn{(2)}}}_{i;jk}(x)-\mu^{{\mbtn{(2)}}}_{j;ik}(x)}{v_k},\vs{6pt}\\

\commut{x^i}{v_j}=i\hbar P_{ij}(x),\vs{6pt}\\

\commut{v_i}{p^x_j}=i\hbar(\symmp{\lambda}{P_{ik}(x)G_{kj}}+\symmp{\mu^{{\mbtn{(2)}}}_{i;jk}(x)}{v_k}),\vs{6pt}\\ 

\commut{x^i}{\lambda}=i\hbar\nu_i(x),\vs{6pt}\\

\commut{\lambda}{p^x_i}=i\hbar(\symmp{\mu^{{\mbtn{(3)}}}_{ik}(x)}{v_k}+\symmp{\lambda}{\mu^{{\mbtn{(2)}}}_{k;ik}(x)}),\vs{6pt}\\

\commut{\lambda}{v_i}=i\hbar\symmp{\mu^{{\mbtn{(3)}}}_{ik}(x)}{v_k},\hs{120pt}(\mbox{the others})=0.

\end{array}
\eqno{(4.19)}
$$
\ts{12pt}Through (4.18) and (4.19), thus, the projected CCS  $\mcC^{\mbtn{(2)}}$ is defined as
$$
\mcC^{\mbtn{(2)}}=\{(x^i,p^x_i),(u_i,p_u^i)|i=1,\cdots,N \}
\eqno{(4.20a)}
$$
with
$$
\begin{array}{l}

v_i=\symmp{P_{ij}(x)}{p^x_j},\hs{12pt}\lambda=-\symmp{\mcG^{=1}(x)G_i(x)}{p^x_i},\vs{12pt}\\

p_v^i=0,\hs{12pt}p_{\lambda}=0,

\end{array}
\eqno{(4.20b)}
$$
which obeys the commutator algebra $\mcA(\mcC^{\mbtn{(2)}})$:
$$
\commut{x^i}{p^x_j}=i\hbar\delta^i_j,\hs{12pt}\commut{u_i}{p^j_u}=i\hbar\delta_i^j,\hs{12pt}\mbox{(the others)}=0.
\eqno{(4.21)}
$$ 
\ts{12pt}From (4.15), the remaining constraints $\phi^{\mbtn{(4)}}_i(i=1,\cdots,N)$ $(\in\mcK^{\mbtn{(C)}})$ are projected on to $\mcC^{\mbtn{(2)}}$ in the following way:
$$
\Phat^{\mbtn{(2)}}\phi^{\mbtn{(4)}}_i=\Phat^{\mbtn{(2)}}(p_u^i+\f12\Theta^{ij}u_j)=p_u^i+\f12\Theta^{ij}u_j=\phi^{\mbtn{(4)}}_i\hs{24pt}\in \mcC^{\mbtn{(2)}}.
\eqno{(4.22)}
$$
Thus,
$$
\mcK^{\mbtn{(C)}}(\in\mcC^{\mbtn{(2)}})=\{\phi^{\mbtn{(4)}}\}\hs{12pt}\mbox{with}\hs{12pt}\mcA(\mcK^{\mbtn{(C)}}):\commut{\phi^{\mbtn{(4)}}_i}{\phi^{\mbtn{(4)}}_j}=i\hbar\Theta^{ij}.
\eqno{(4.23)}
$$

\subsubsection{Projected Hamiltonian $H^{\mbtn{(2)}}$}

\ts{12pt}With the similar procedure in $\mcS^{\mbtn{(1)}}$, $H^{\mbtn{(2)}}$ is obtained in  the following way:
$$
H^{\mbtn{(2)}}=\PhatB^{\mbtn{(2)}}\hat{{\bf I}}^{\mbtn{(2)}}H^{\mbtn{(1)}}=\Phat^{\mbtn{(2)}}H^{\mbtn{(1)}}+\Qhat^{\mbtn{(2)}}H^{\mbtn{(1)}},
\eqno{(4.24a)}
$$
where
$$
\Qhat^{\mbtn{(2)}}=\sum^{\infty}_{n+m\neq 0}\f1{n!m!}\left(\f{\hbar}4\right)^{n+m}\Phat^{\mbtn{(2)}}(\pimm{(2)})^{2m}(\ximm{(2)})^{2n}.
\eqno{(4.24b)}
$$
\vs{12pt}\\
{\bf i)} $\Phat^{\mbtn{(2)}}H^{\mbtn{(1)}}$\vs{12pt}\\
\ts{12pt}Since $H^{\mbtn{(1)}}$ is rewritten as 
$$
H^{\mbtn{(1)}}=\dps{-\f{\hbar}4N+\f12v_iv_i}+U_{\mbtn{I}}(x)+\symmp{U_{\mbtn{II}}^{ij}(x)}{v_iv_j}
+\pipp{(2)}U_{\mbtn{III}}(x)+\pipp{(2)}\symmp{U_{\mbtn{IV}}^{ij}(x)}{v_iv_j},
\eqno{(4.25)}
$$
then, $\Phat^{\mbtn{(2)}}H^{\mbtn{(1)}}$ becomes as
$$
\begin{array}{l}

\Phat^{\mbtn{(2)}}H^{\mbtn{(1)}}\vs{12pt}\\

\dps{=-\f{\hbar}4N+\f12\symmp{\delta_{ij}+2U_{\mbtn{II}}^{ij}(x)}{\symmp{v_i}{v_j}}}+\dps{\f12U^v_{ii}(x)+U_{\mbtn{I}}(x)+U^v_{ij}(x)U_{\mbtn{II}}^{ij}(x)}\vs{12pt}\\

-\dps{\f{\hbar^2}4G_i(x)G_j(x)(\nu(x)\partial)^2U_{\mbtn{II}}^{ij}(x)+\f{\hbar^2}2G_i(x)P_{jk}(x)((\nu(x)\partial)U_{\mbtn{II}}^{ij}(x))_{;k}},
\end{array}
\eqno{(4.26a)}
$$
where $(\nu(x)\partial)=(\nu_i(x)\partial^x_i)$ and $U^v_{ij}(x)$ is the quantum correction due to the projection of $\symmp{v_i}{v_j}$:
$$
U^v_{ij}(x)=\f{\hbar^2}2\mu^{{\mbtn{(3)}}}_{ik}(x)G_{kj}(x)+\f{3\hbar^2}4\mu^{{\mbtn{(2)}}}_{k;ik}(x)\mu^{{\mbtn{(2)}}}_{l;jl}(x).
\eqno{(4.26b)}
$$
\vs{12pt}\\
{\bf ii)} $\Qhat^{\mbtn{(2)}}H^{\mbtn{(1)}}$\vs{12pt}\\
$\Qhat^{\mbtn{(2)}}H^{\mbtn{(1)}}$ is represented with
$$
\begin{array}{lcl}

\Qhat^{\mbtn{(2)}}H^{\mbtn{(1)}}&=&\Qhat^{\mbtn{(2)}}\dps{\f12}\symmp{v_i}{v_i}+\Qhat^{\mbtn{(2)}}U_{\mbtn{I}}(x)+\Qhat^{\mbtn{(2)}}\symmp{U_{\mbtn{II}}^{ij}(x)}{v_iv_j}\vs{12pt}\\

&+&\Qhat^{\mbtn{(2)}}\symmp{U_{\mbtn{III}}(x)}{p_{\lambda}}+\Qhat^{\mbtn{(2)}}\symmp{\symmp{U_{\mbtn{IV}}^{ij}(x)}{v_iv_j}}{p_{\lambda}},

\end{array}
\eqno{(4.27)}
$$
the explisit form of which is presented in Appendix B.
\vs{12pt}\\
iii) Projected Hamiltonian $H^{\mbtn{(2)}}$
\vs{12pt}\\
\ts{12pt}From (4.26a) and (4.27), the projected Hamiltonian $H^{\mbtn{(2)}}$ is represented in the following form:
$$
\begin{array}{lcl}

H^{\mbtn{(2)}}&=&\Phat^{\mbtn{(2)}}H^{\mbtn{(1)}}+\Qhat^{\mbtn{(2)}}H^{\mbtn{(1)}}\vs{12pt}\\

&=&-\dps{\f{\hbar}4}N+\dps{\f12}\symmp{P_{ik}(x)M^{\mbtn{(2)}}_{kl}(x)P_{lj}(x)}{\symmp{p^x_i}{p^x_j}}+U^{\mbtn{(2)}}_{\mbtn{I}}(x)+U^{\mbtn{(2)}}_{\mbtn{II}}(x),

\end{array}
\eqno{(4.28a)}
$$
where
$$
\begin{array}{lcl}

M^{\mbtn{(2)}}_{ij}(x)&=&\delta_{ij}+m^{\mbtn{(2)}}_{ij}(x),\vs{12pt}\\

m^{\mbtn{(2)}}_{ij}(x)&=&2U_{\mbtn{(II)}}^{ij}(x)+B^{ij}(x)+2\mcM_{\mbtn{(II)}}^{ij}(x)+\hbar\mcM_{\mbtn{(IV)}}^{ij}(x),

\end{array}
\eqno{(4.28b)}
$$
and, $B^{ij}(x)$, $\mcM_{\mbtn{(II)}}^{ij}(x)$ and $\mcM_{\mbtn{(IV)}}^{ij}(x)$ are presented in Appendix B. 
\vs{6pt}\\
\ts{12pt}Here, $U^{\mbtn{(2)}}_{\mbtn{I}}(x)$ is the quantum correction due to the operator re-ordering\\
 in $\dps{\f12}\symmp{M^{\mbtn{(2)}}_{ij}(x)}{\symmp{v_i}{v_j}}=\dps{\f12}\symmp{M^{\mbtn{(2)}}_{ij}(x)}{\symmp{\symmp{P_{ik}(x)}{p^x_k}}{\symmp{P_{jl}(x)}{p^x_l}}}$:

$$
\begin{array}{lcl}

U^{\mbtn{(2)}}_{\mbtn{I}}(x)&=&\dps{\f{\hbar^2}8}M^{\mbtn{(2)}}_{ij}(x)((P_{ik}(x)P_{jl}(x))_{;kl}+P_{ik}(x)_{;l}P_{jl}(x)_{;k})\vs{12pt}\\

&+&\dps{\f{\hbar^2}8}(M^{\mbtn{(2)}}_{ij}(x)_{;k}(P_{ik}(x)P_{jl}(x))_{;l}+(P_{ik}(x)P_{jl}(x))_{;k}M^{\mbtn{(2)}}_{ij}(x)_{;l},

\end{array}
\eqno{(4.29a)}
$$
and $U^{\mbtn{(2)}}_{\mbtn{II}}(x)$, the quanum correction associated to the ACCS expansion with $\Phat^{\mbtn{(2)}}$:
$$
\begin{array}{lcl}

U^{\mbtn{(2)}}_{\mbtn{II}}(x)
&=&\dps{\f12}M^{\mbtn{(2)}}_{ij}(x)U^v_{ij}(x)+\dps{\f{\hbar^2}8}\mcB_{ii}(x)

+\dps{\sum^{\infty}_{n= 0}\f1{n!}\left(\f{\hbar}4\right)^n(\nu(x)\partial)^{2n}(\f{\hbar}4\mcG(x)+U_{\mbtn{I}}(x))}\vs{12pt}\\

&-&\dps{\f{\hbar^2}8G_i(x)G_j(x)(\nu(x)\partial)^2m^{\mbtn{(2)}}_{ij}(x)+\f{\hbar^2}4G_i(x)P_{jk}(x)(\nu(x)\partial)m^{\mbtn{(2)}}_{ij}(x)_{;k}}\vs{12pt}\\

&+&\dps{\f{\hbar}2\sum^{\infty}_{n= 0}\f1{n!}\left(\f{\hbar}4\right)^n(\nu(x)\partial)^{2n}(U_{\mbtn{II}}^{ij}(x)G_i(x)G_j(x)+\nu_k(x)U_{\mbtn{III}}(x)_{;k})}\vs{12pt}\\

&+&\dps{\f{\hbar^2}4\sum^{\infty}_{n= 0}\f1{n!}\left(\f{\hbar}4\right)^n(\nu(x)\partial)^{2n}(\nu_k(x)(U_{\mbtn{IV}}^{ij}(x)G_i(x)G_j(x))_{;k})}+\mcU^{\mbtn{(II)}}(x)+\mcU^{\mbtn{(IV)}}(x),

\end{array}
\eqno{(4.29b)}
$$ 
where $\mcB_{ii}(x)$ is presented in Appendix B.1, and $\mcU^{\mbtn{(II)}}(x)$ is defined in Appendix B.3, $\mcU^{\mbtn{(IV)}}(x)$, defined in Appendix B.5.
\vs{6pt}\\
\ts{12pt}Thus, we have constructed $\mcS^{\mbtn{(2)}}=(\mcC^{\mbtn{(2)}}, H^{\mbtn{(2)}},\mcK^{\mbtn{(C)}})$.

\subsection{Construction of $\mcS^{\mbtn{(3)}}$}

\ts{12pt}In order to eliminate the remaining constraint-set $\mcK^{(\mbtn{C})}$, we shall impose the additional constraints, which produce the noncommutativity for the CCS $\mcC^{\mbtn{(2)}}$. \\
\ts{12pt}For this purpose, here, we shall prepare the following quantities:
$$
\begin{array}{l}

\Xi_{ij}=\eta\varepsilon^{ij}\q(\eta:\mbox{the constant parameter}),\vs{6pt}\\

G_{ij}=(\Theta\Xi)_{ij}=(\Xi\Theta)_{ij}\hs{12pt}(i,j=1,\cdots,N),\vs{6pt}\\

M_{ij}=(I+\dps{\f14}G)_{ij}=\delta_{ij}+\dps{\f14}G_{ij},\vs{6pt}\\

\bM_{ij}=(I-\dps{\f14}G)_{ij}=\delta_{ij}-\dps{\f14}G_{ij}.
\end{array}
\eqno{(4.30)}
$$
 
\subsubsection{ Additional Constraints}
 
\ts{12pt}Following our previous works\cite{S2}, we shall impose the additional constraints $\psi^{\mbtn{(2)}}_i$ ($i=1,\cdots,N)$):
$$
\psi^{\mbtn{(2)}}_i=\bM_{ij}u_j-p^x_i-\f12\Xi_{ij}x^j\hs{36pt}(i=1,\cdots,N),
 \eqno{(4.31)}
$$
\vs{6pt}\\
and therefore, the constraint-set on $\mcC^{\mbtn{(2)}}$ is defined as   
$$
\mcK^{(\mbtn{C$^*$})}=\{\phi^{\mbtn{(4)}}_i,\psi^{\mbtn{(2)}}_i|i=1,\cdots,N\}
\eqno{(4.32a)}
$$
with $\mcA(\mcK^{(\mbtn{C$^*$})})$:
$$
\commut{\phi^{\mbtn{(4)}}_i}{\phi^{\mbtn{(4)}}_j}=i\hbar\Theta^{ij},\hs{12pt}\commut{\phi^{\mbtn{(4)}}_i}{\psi^{\mbtn{(3)}}_j}=i\hbar\bM_{ij},\hs{12pt}\commut{\psi^{\mbtn{(3)}}_i}{\psi^{\mbtn{(3)}}_j}=i\hbar\Xi_{ij}.
\eqno{(4.32b)}
$$
Then, the ACCS associated with $\mcK^{(\mbtn{C$^*$})}$ and the {\it hyper}-operator $\Qhat^{\mbtn{(3)}}_{\eta\zeta}$  are defined, respectively, as follows:
$$
Z^{\mbtn{(3)}}_{\alpha}=\left\{\begin{array}{lcll}\xi^{\mbtn{(3)}}_i&=&M^{-1}_{ij}(\psi^{\mbtn{(2)}}_j+\dps{\f12}\Xi_{jk}\phi^{\mbtn{(4)}}_k)\vs{6pt}\\

&=&M^{-1}_{ij}(u_j+\dps{\f12}\Xi_{jk}p_u^k-p^x_j-\dps{\f12}\Xi_{jk}x^k)&\q (\alpha=i)\vs{12pt}\\

\pi^{\mbtn{(3)}}_i&=&M^{-1}_{ij}(\phi^{\mbtn{(4)}}_j-\dps{\f12}\Theta_{jk}\psi^{\mbtn{(2)}}_k)\vs{6pt}\\

&=&M^{-1}_{ij}(p_u^j+\dps{\f18}(G\Theta)_{jk}u_k+\dps{\f12}\Theta^{jk}p^x_k+\dps{\f14}G_{jk}x^k)&\q (\alpha=i+N),\end{array}\right.
\eqno{(4.33)}
$$ 
and
$$
\Qhat^{\mbtn{(3)}}_{\eta\zeta}=J^{\alpha\beta}\Zmms{\mbtn{(3)}}{\alpha}(\eta)\Zmms{\mbtn{(3)}}{\beta}(\zeta)=\hxi^{\mbtn{(3)}(-)}_i(\eta)\hpi^{\mbtn{(3)}(-)}_i(\zeta)-\hpi^{\mbtn{(3)}(-)}_i(\eta)\hxi^{\mbtn{(3)}(-)}_i(\zeta)
\eqno{(4.34a)}
$$
with
$$
\begin{array}{lclclcl}

\hxi^{\mbtn{(3)}(-)}_kx^i&=&M^{-1}_{ki},& &\hpi^{\mbtn{(3)}(-)}_kx^i&=&-\dps{\f12}(M^{-1}\Theta)_{ki},\vs{6pt}\\

\hxi^{\mbtn{(3)}(-)}_kp^x_i&=&-\dps{\f12}(M^{-1}\Xi)_{ki},& &\hpi^{\mbtn{(3)}(-)}_kp^x_i&=&\dps{\f14}(M^{-1}G)_{ki},\vs{6pt}\\

\hxi^{\mbtn{(3)}(-)}_ku_i&=&-\dps{\f12}(M^{-1}\Xi)_{ki},& &\hpi^{\mbtn{(3)}(-)}_ku_i&=&-M^{-1}_{ki}.

\end{array}
\eqno{(4.34b)}
$$
 
\subsubsection{The projected CCS $\mcC^{\mbtn{(3)}}$} 

\ts{12pt}The projected CCS  $\mcC^{\mbtn{(3)}}$ is obtained as follows\cite{S2}:
$$
\mcC^{\mbtn{(3)}}=\Phat^{\mbtn{(3)}}\mcC^{\mbtn{(2)}}=\{(x^i,p^x_i),(u_i,p_u^i)|i=1,\cdots,N\}
\eqno{(4.35a)}
$$
with
$$
\begin{array}{lcl}

p_u^i+\dps{\f12}\Theta^{ij}p^x_j=0,& &\bM_{ij}u_j-p^x_i-\dps{\f12}\Xi_{ij}x^i=0,\vs{12pt}\\

v_i=\symmp{P_{ij}(x)^*}{p^x_j},&  &p_v^i=0,\vs{12pt}\\

\lambda=-\symmp{(\mcG^{-1}(x)G_i(x))^*}{p^x_i},& &p_{\lambda}=0.

\end{array}
\eqno{(4.35b)}
$$
where $\Phat^{\mbtn{(3)}}$ is the projection operator for $\mcK^{(\mbtn{C$^*$})}$, which satisfies 
$$
\Phat^{\mbtn{(3)}}\mcK^{(\mbtn{C$^*$})}=0,
\eqno{(4.35c)}
$$
and, for any operator $O(x)$, $O(x)^*=\Phat^{\mbtn{(3)}}O(x)$.  
Then, the commutator algebra $\mcA(\mcC^{\mbtn{(3)}})$ is given by
$$
\begin{array}{lcl}

\commut{x^i}{x^j}=i\hbar(M^{-1}\Theta M^{-1})_{ij},&  &\commut{u_i}{u_j}=i\hbar(M^{-1}\Xi M^{-1})_{ij},\vs{12pt}\\

\commut{x^i}{p^x_j}=i\hbar(M^{-1}(I+\dps{\f1{16}}G^2)M^{-1})_{ij},&  &\commut{u_i}{p_u^j}=i\hbar\dps{\f12}(M^{-1}GM^{-1})_{ij},\vs{12pt}\\

\commut{p^x_i}{p^x_j}=-i\hbar\dps{\f14}(M^{-1}G\Xi M^{-1})_{ij},&  &\commut{p_u^i}{p_u^j}=-i\hbar\dps{\f14}(M^{-1}G\Theta M^{-1})_{ij},\vs{12pt}\\

\commut{x^i}{u_j}=i\hbar(M^{-1}\bM M^{-1})_{ij},& &\commut{u_i}{p^x_j}=i\hbar\dps{\f12}(M^{-1}\Xi\bM M^{-1})_{ij},\vs{12pt}\\

\commut{x^i}{p_u^j}=i\hbar\dps{\f12}(M^{-1}\Theta\bM M^{-1})_{ij},& &\commut{p^x_i}{p_u^j}=i\hbar\dps{\f14}(M^{-1}G\bM M^{-1})_{ij}.

\end{array}
\eqno{(4.36)}
$$ 

\subsubsection{Projected Hamiltonian $H^{\mbtn{(3)}}$}

\ts{12pt}As well as in $\mcS^{\mbtn{(1)}}$ and $\mcS^{\mbtn{(2)}}$, the projected Hamiltonian $H^{\mbtn{(3)}}$ in $\mcS^{\mbtn{(3)}}$ is obtained in the following way:
$$
H^{\mbtn{(3)}}=\PhatB^{\mbtn{(3)}}\hat{{\bf I}}^{\mbtn{(3)}}H^{\mbtn{(2)}}=\Phat^{\mbtn{(3)}}H^{\mbtn{(2)}}+\Qhat^{\mbtn{(3)}}H^{\mbtn{(2)}},
\eqno{(4.37a)}
$$
where
$$
\Qhat^{\mbtn{(3)}}=\dps{\sum^{\infty}_{n+m\neq 0}\f1{n!m!}\left(\f{\hbar}4\right)^{n+m}\Phat^{\mbtn{(3)}}(\pimm{(3)}_k\pimm{(3)}_k)^m(\ximm{(3)}_l\ximm{(3)}_l)^n}. 
\eqno{(4.37b)}
$$
For the simplicity, here, we shall express $H^{\mbtn{(2)}}$ with
$$
H^{\mbtn{(2)}}=-\dps{\f{\hbar}4}N+\dps{\f12}\symmp{\tM^{\mbtn{(2)}}_{ij}(x)}{\symmp{p^x_i}{p^x_j}}+U^{\mbtn{(2)}}(x),
\eqno{(4.38a)}
$$
where
$$
\begin{array}{l}
\tilde{M}^{\mbtn{(2)}}_{ij}(x)=P_{ik}(x)M^{\mbtn{(2)}}_{kl}(x)P_{lj}(x),\vs{6pt}\\
U^{\mbtn{(2)}}(x)=U^{\mbtn{(2)}}_{\mbtn{I}}(x)+U^{\mbtn{(2)}}_{\mbtn{II}}(x).

\end{array}
\eqno{(4.38b)}
$$
\vs{6pt}\\
{\boldmath 1) $\Phat^{\mbtn{(3)}}H^{\mbtn{(2)}}$}
\vs{6pt}\\
\ts{12pt}The projected term $\Phat^{\mbtn{(3)}}H^{\mbtn{(2)}}$ is obtained in the following way\footnote{For any operator $O(x)$, $O(x)^*=\Phat^{\mbtn{(3)}}O(x)$}:
$$
\begin{array}{lcl}

\Phat^{\mbtn{(3)}}H^{\mbtn{(2)}}&=&-\dps{\f{\hbar}4}N+\Phat^{\mbtn{(3)}}\dps{\f12}\symmp{\tilde{M}^{\mbtn{(2)}}_{ij}(x)}{\symmp{p^x_i}{p^x_j}}+\Phat^{\mbtn{(3)}}U^{\mbtn{(2)}}(x)\vs{12pt}\\

&=&-\dps{\f{\hbar}4}N+\dps{\f12}\symmp{\tM^{\mbtn{(2)}}_{ij}(x)^*}{\symmp{p^x_i}{p^x_j}}+U^{\mbtn{(3)}}_{\mbtn{I}}(x)+U^{\mbtn{(3)}}_{\mbtn{II}}(x),

\end{array}
\eqno{(4.39)}
$$
where
$$
\begin{array}{lcl}

U^{\mbtn{(3)}}_{\mbtn{I}}(x)&=&-\dps{\f{\hbar^2}{16}}\left((M^{-1}GM^{-1})_{ik}\tM^{\mbtn{(2)}}_{ij}(x)^*_{;kj}+(M^{-1}GM^{-1})_{jk}\tM^{\mbtn{(2)}}_{ij}(x)^*_{;ki}\right)\vs{12pt}\\

&+&\dps{\f{\hbar^2}{32}}(M^{-1}GM^{-1})_{ik}(M^{-1}GM^{-1})_{jl}\tM^{\mbtn{(2)}}_{ij}(x)^*_{;kl},

\end{array}
\eqno{(4.40a)}
$$
which is the quantum correction associated to the projedtion of $\dps{\f12}\symmp{\tilde{M}^{\mbtn{(2)}}_{ij}(x)}{\symmp{p^x_i}{p^x_j}}$,\\
and 
$$
U^{\mbtn{(3)}}_{\mbtn{II}}(x)=U^{\mbtn{(2)}}(x)^*=U^{\mbtn{(2)}}_{\mbtn{I}}(x)^*+U^{\mbtn{(2)}}_{\mbtn{II}}(x)^*.
\eqno{(4.40b)}
$$ 
\vs{6pt}\\
{\boldmath 2) $\Qhat^{\mbtn{(3)}}H^{\mbtn{(2)}}$}
\vs{6pt}\\
\ts{12pt}The {\it ACCS}-expansion term $\Qhat^{\mbtn{(3)}}H^{\mbtn{(2)}}$ is defined as
$$
\Qhat^{\mbtn{(3)}}H^{\mbtn{(2)}}=\f12\Qhat^{\mbtn{(3)}}\symmp{\tilde{M}^{\mbtn{(2)}}_{ij}(x)}{\symmp{p^x_i}{p^x_j}}+\Qhat^{\mbtn{(3)}}U^{\mbtn{(2)}}(x).
\eqno{(4.41)}
$$
From (4.34b), then, $\Qhat^{\mbtn{(3)}}U^{\mbtn{(2)}}(x)$ is obtained as follows:
$$
\begin{array}{l}

\Qhat^{\mbtn{(3)}}U^{\mbtn{(2)}}(x)=\dps{\sum^{\infty}_{n+m\neq 0}\f1{n!m!}\left(\f{\hbar}4\right)^{n+m}\Phat^{\mbtn{(3)}}(\hpi^{\mbtn{(3)}(-)}_{l_m})^2\cdots(\hpi^{\mbtn{(3)}(-)}_{l_1})^2(\hxi^{\mbtn{(3)}(-)}_{k_n})^2\cdots(\hxi^{\mbtn{(3)}(-)}_{k_1})^2U^{\mbtn{(2)}}(x)}\vs{12pt}\\

=\dps{\sum^{\infty}_{n+m\neq 0}\f1{n!m!}\left(\f{\hbar}4\right)^{n+m}\left(\f14\right)^m}\vs{12pt}\\

\times(M^{-1}\Theta^2M^{-1})_{l_{2m}l_{2m-1}}\cdots(M^{-1}\Theta^2M^{-1})_{l_2l_1}(M^{-2})_{k_{2n}k_{2n-1}}\cdots(M^{-2})_{k_2k_1}U^{\mbtn{(2)}}(x)^*_{;k_1\cdots k_{2n},l_1\cdots l_{2m}}.

\end{array}
\eqno{(4.42)}
$$
Through the tedious calculations,  $\dps{\f12}\Qhat^{\mbtn{(3)}}\symmp{\tilde{M}^{\mbtn{(2)}}_{ij}(x)}{\symmp{p^x_i}{p^x_j}}$ is obtained in the following way:
$$
\begin{array}{l}

\dps{\f12}\Qhat^{\mbtn{(3)}}\symmp{\tM^{\mbtn{(2)}}_{ij}(x)}{\symmp{p^x_i}{p^x_j}}\vs{12pt}\\

=\dps{\f12}\dps{\sum^{\infty}_{n+m\neq 0}\f1{n!m!}\left(\f{\hbar}4\right)^{n+m}\Phat^{\mbtn{(3)}}(\hpi^{\mbtn{(3)}(-)}_{b_m})^2\cdots(\hpi^{\mbtn{(3)}(-)}_{b_1})^2(\hxi^{\mbtn{(3)}(-)}_{a_n})^2\cdots(\hxi^{\mbtn{(3)}(-)}_{a_1})^2\symmp{\tM^{\mbtn{(2)}}_{ij}(x)}{\symmp{p^x_i}{p^x_j}}}\vs{12pt}\\

=\dps{\f12}\symmp{X_{ij}(x)^*}{\symmp{p^x_i}{p^x_j}}\vs{12pt}\\

-\dps{\f{\hbar}{16}}(M^{-1}G\Theta M^{-1})_{ik}\symmp{\tX_{ij}(x)^*_{;k}}{p^x_j}+\dps{\f{\hbar}4}(M^{-1}\Xi  M^{-1})_{ik}\symmp{\tX_{ij}(x)^*_{;k}}{p^x_j}\vs{12pt}\\

+U^{\mbtn{(3)}}_{\mbtn{Q}}(x),

\end{array}
\eqno{(4.43)}
$$
where
$$
\begin{array}{l}

U^{\mbtn{(3)}}_{\mbtn{Q}}
=\dps{\f{\hbar}{2\cdot 32}}(M^{-1}G^2M^{-1})_{ij}\tX_{ij}(x)^*-\dps{\f{\hbar}{2\cdot8}}(M^{-1}\Xi^2M^{-1})_{ij}\tX_{ij}(x)^*\vs{12pt}\\

+\dps{\f{\hbar^2}{2\cdot16^2}}(M^{-1}G\Theta M^{-1})_{ik}(M^{-1}G\Theta  M^{-1})_{jl}\tX_{ij}(x)^*_{;kl}+\dps{\f{\hbar^2}{2\cdot16}}(M^{-1}\Xi  M^{-1})_{ik}(M^{-1}\Xi M^{-1})_{jl}\tX_{ij}(x)^*_{;kl}\vs{12pt}\\

-\dps{\f{\hbar^2}{2\cdot32}}(M^{-1}\Xi  M^{-1})_{ik}(M^{-1}G\Theta M^{-1})_{jl}\tX_{ij}(x)_{;kl}^*\vs{12pt}\\

-\dps{\f{\hbar^2}{2\cdot4}}(M^{-1}GM^{-1})_{ik}X_{ij}(x)^*_{;kj}+\dps{\f{\hbar^2}{2\cdot16}}(M^{-1}GM^{-1})_{ik}(M^{-1}GM^{-1})_{jl}X_{ij}(x)^*_{;kl},

\end{array}
\eqno{(4.44a)}
$$
and
$$
\begin{array}{l}

X_{ij}(x)=\dps{\sum^{\infty}_{n+m\neq 0}\f1{n!m!}\left(\f{\hbar}4\right)^{n+m}}X^{(m,n)}_{ij}(x),
\vs{12pt}\\
\tX_{ij}(x)=X^{(0,0)}_{ij}(x)+X_{ij}(x)=\dps{\sum^{\infty}_{n,m= 0}\f1{n!m!}\left(\f{\hbar}4\right)^{n+m}}X^{(m,n)}_{ij}(x)

\end{array}
\eqno{(4.44b)}
$$
with
$$
\begin{array}{l}
X^{(m,n)}_{ij}(x)=(\hpi^{\mbtn{(3)}(-)}_{b_m})^2\cdots(\hpi^{\mbtn{(3)}(-)}_{b_1})^2(\hxi^{\mbtn{(3)}(-)}_{a_n})^2\cdots(\hxi^{\mbtn{(3)}(-)}_{a_1})^2\tM^{\mbtn{(2)}}_{ij}(x).

\end{array}
\eqno{(4.44c)}
$$
\vs{6pt}\\
{\boldmath 3) Projected Hamiltonian $H^{\mbtn{(3)}}$}\vs{6pt}\\
\ts{12pt}From (4.37a), (4.39) and (4.43), consequently, the projected Hamiltonian $H^{\mbtn{(3)}}$ is given  in the following way:
$$
\begin{array}{l}

H^{\mbtn{(3)}}=\PhatB^{\mbtn{(3)}}\hat{{\bf I}}^{\mbtn{(3)}}H^{\mbtn{(2)}}=\Phat^{\mbtn{(3)}}H^{\mbtn{(2)}}+\Qhat^{\mbtn{(3)}}H^{\mbtn{(2)}}\vs{12pt}\\

=-\dps{\f{\hbar}4}N+\dps{\f12}\symmp{\tM^{\mbtn{(2)}}_{ij}(x)^*}{\symmp{p^x_i}{p^x_j}}+U^{\mbtn{(3)}}_{\mbtn{I}}(x)+U^{\mbtn{(3)}}_{\mbtn{II}}(x)\vs{12pt}\\

+\dps{\f12}\symmp{X_{ij}(x)^*}{\symmp{p^x_i}{p^x_j}}\vs{12pt}\\

-\dps{\f{\hbar}{16}}(M^{-1}G\Theta M^{-1})_{ik}\symmp{\tX_{ij}(x)^*_{;k}}{p^x_j}+\dps{\f{\hbar}4}(M^{-1}\Xi  M^{-1})_{ik}\symmp{\tX_{ij}(x)^*_{;k}}{p^x_j}\vs{12pt}\\

+U^{\mbtn{(3)}}_{\mbtn{Q}}(x)\vs{12pt}\\

=-\dps{\f{\hbar}4}N\vs{12pt}\\

+\dps{\f12}\symmp{\tX_{ij}(x)^*}{\symmp{p^x_i}{p^x_j}}

+\dps{\f{\hbar}4}\symmp{{\mathcal M}_{kl}\tX_{ik}(x)^*_{;l}}{p^x_i}\vs{12pt}\\

+U^{\mbtn{(3)}}_{\mbtn{I}}(x)+U^{\mbtn{(3)}}_{\mbtn{II}}(x)+U^{\mbtn{(3)}}_{\mbtn{Q}}(x),

\end{array}
\eqno{(4.45a)}
$$
where
$$
{\mathcal M}=M^{-1}\Xi(I-\dps{\f14}\Theta^2)M^{-1}.
\eqno{(4.45b)}
$$
\ts{12pt}Through the sequential projections for $\mcK$, thus, we have obtained the final projected system
$$
\mcS^{\mbtn{(3)}}=(\mcC^{\mbtn{(3)}},H^{\mbtn{(3)}},\mcK=0),
\eqno{(4.46)}
$$ 
where $\mcS^{\mbtn{(3)}}$ contains the quantum corrections associated to the constraints $\mcK=0$, which have never appeared in the previous approaches.

\subsection{Constraint Quantum System $\mcS^*$}

\ts{12pt}Taking account of the projection conditions (4.35b) and the commutator algebra $\mcA(\mcC^{\mbtn{(3)}})$, (4.36), in $\mcS^{\mbtn{(3)}}$,  two kinds of constraint quantum system are defined for $\mcS^*$, which we shall denote with $\mcS^*_{\mbtn{I}}$ and $\mcS^*_{\mbtn{II}}$, respectively.

\subsubsection{Constraint Quantum System $\mcS^*_{\mbtn{I}}$}
  
The constraint quantum system  $\mcS^*_{\mbtn{I}}$ is defined with the CCS $\{(x^i,p^x_i)|i=1,\cdots,N\}$ in the following way:
$$
\mcS^*_{\mbtn{I}}=(\mcC^*_{\mbtn{I}},H^*_{\mbtn{I}}).
\eqno{(4.47)}
$$
Here,
$$
\mcC^*_{\mbtn{I}}=\{(x^i,p^x_i|i=1,\cdots,N\}
\eqno{(4.48a)}
$$
with
$$
\begin{array}{lcl}

u_i=\bM^{-1}_{ij}(p^x_j+\dps{\f12}\Xi_{jk}x^k),&  &p_u^i=-\dps{\f12}\Theta^{ij}u_j=-\dps{\f12}(\Theta\bM^{-1})_{ij}(p^x_j+\dps{\f12}\Xi_{jk}x^k),\vs{12pt}\\

v_i=\symmp{P_{ij}(x)^*}{p^x_j},&  &p_v^i=0,\vs{12pt}\\

\lambda=-\symmp{(\mcG(x)^{-1}G_i(x))^*}{p^x_j},& &p_{\lambda}=0,

\end{array}
\eqno{(4.48b)}
$$
which obeys the commutator algebra $\mcA(\mcC^*_{\mbtn{I}})$:
$$
\begin{array}{ll}

\mcA(\mcC^*_{\mbtn{I}}):

&\commut{x^i}{x^j}=i\hbar(M^{-1}\Theta M^{-1})_{ij},\vs{12pt}\\

&\commut{x^i}{p^x_j}=i\hbar(M^{-1}(\dps{I+\f1{16}G^2})M^{-1})_{ij},\vs{12pt}\\

&\commut{p^x_i}{p^x_j}=-i\hbar\dps{\f14}(M^{-1}G\Xi M^{-1})_{ij}.

\end{array}
\eqno{(4.49)}
$$
Then, the resultant Hamiltonian $H^*_{\mbtn{I}}$ becomes as follows:
$$
H^*_{\mbtn{I}}=H^{(3)}(x,p^x)=H^{(3)}.
\eqno{(4.50)}
$$

\subsubsection{Constraint Quantum System $\mcS^*_{\mbtn{II}}$} 

\ts{12pt}As well as in $\mcS^*_{\mbtn{I}}$, the constraint quantum system $\mcS^*_{\mbtn{II}}$ is constructed with the CCS $\{(x^i,u_i)|i=1,\cdots,N\}$ in the following way: 
$$
\mcS^*_{\mbtn{II}}=(\mcC^*_{\mbtn{II}},H^*_{\mbtn{II}}),
\eqno{(4.51)}
$$
where,
$$
\mcC^*_{\mbtn{II}}=\{(x^i,u_i)|i=1,\cdots,N\}
\eqno{(4.52a)}
$$
with
$$
\begin{array}{lcl}

p^x_i=\bM_{ij}u_j-\dps{\f12}\Xi_{ij}x^j,&  &p_u^i=-\dps{\f12}\Theta^{ij}u_j,\vs{12pt}\\

v_i=\symmp{P_{ij}(x)^*}{\bM_{jk}u_k-\dps{\f12}\Xi_{jk}x^k},&  &p_v^i=0,\vs{12pt}\\

\lambda=-\symmp{(\mcG(x)^{-1}G_i(x))^*}{\bM_{jk}u_k-\dps{\f12}\Xi_{jk}x^k},& &p_{\lambda}=0,

\end{array}
\eqno{(4.52b)}
$$
of which the commutator algebra $\mcA(\mcC^*_{\mbtn{II}})$ is defined by
$$
\begin{array}{ll}

\mcA(\mcC^*_{\mbtn{II}}):

&\commut{x^i}{x^j}=i\hbar(M^{-1}\Theta M^{-1})_{ij},\vs{12pt}\\

&\commut{x^i}{u_j}=i\hbar(M^{-1}\bM M^{-1})_{ij},\vs{12pt}\\

&\commut{u_i}{u_j}=i\hbar(M^{-1}\Xi M^{-1})_{ij}.

\end{array}
\eqno{(4.53)}
$$
From the projection conditions (4.52b), the resultant Hamiltonian $H^*_{\mbtn{II}}$ is obtained in the  following way:
$$
\begin{array}{l}

H^*_{\mbtn{II}}=H^{(3)}(x,p^x(x,u))\vs{12pt}\\

=-\dps{\f{\hbar}4}N\vs{12pt}\\

+\dps{\f12}\symmp{\tX_{ij}(x)^*}{\symmp{p^x_i(x,u)}{p^x_j(x,u)}}

+\dps{\f{\hbar}4}\symmp{{\mathcal M}_{kl}\tX_{ik}(x)^*_{;l}}{p^x_i(x,u)}\vs{12pt}\\

+U^{\mbtn{(3)}}_{\mbtn{I}}(x)+U^{\mbtn{(3)}}_{\mbtn{II}}(x)

+U^{\mbtn{(3)}}_{\mbtn{Q}}(x)\vs{12pt}\\

=-\dps{\f{\hbar}4}N\vs{12pt}\\

+\dps{\f12}\symmp{\bM_{ik}\tX_{kl}(x)^*\bM_{lj}}{\symmp{u_i}{u_j}}+A_{\mbtn{K}}(x,u)\vs{12pt}\\

+U^{\mbtn{(3)}}_{\mbtn{I}}(x)+U^{\mbtn{(3)}}_{\mbtn{II}}(x)

+U^{\mbtn{(3)}}_{\mbtn{Q}}(x),

\end{array}
\eqno{(4.54a)}
$$
where
$$
\begin{array}{l}

A_{\mbtn{K}}(x,u)=\dps{\f12}\symmp{\Xi_{ik}\tX_{kl}(x)^*\bM_{lj}}{\symmp{x^i}{u_j}}

-\dps{\f18}\symmp{\Xi_{ik}\tX_{kl}(x)^*\Xi_{lj}}{\symmp{x^i}{x^j}}\vs{12pt}\\

+\dps{\f{\hbar}4}\symmp{\bM_{ij}\tX_{jk}(x)^*_{;l}\mathcal{M}_{kl}}{u_i}+\dps{\f{\hbar}8}\symmp{\Xi_{ij}\tX_{jk}(x)^*_{;l}\mathcal{M}_{kl}}{x^i},

\end{array}
\eqno{(4.54b)}
$$
which is the additional term caused by representing $p^x$ in terms of $u$.\vs{6pt}\\

Thus, we have constructed the constraint quantum system $\mcS^*_{\mbtn{II}}$:
$$
\mcS^*_{\mbtn{II}}=(\mcC^*_{\mbtn{II}}(x,u),H^*_{\mbtn{II}}(x,u)).
\eqno{(4.55)}
$$
 
\section{Discussion and Concluding remarks}
\ts{12pt}In order to construct the noncommutative quantum system on the curved space {\it exactly}, we have proposed the Lagrangian $L=L(x,\dot{x},v,\dot{v},\lambda,\dot{\lambda},u,\dot{u})$, (3.2a), with the {\it dynamical} constraint, which has been obtained by modifying the first-order singular Lagrangians in noncommutative quantum theories.\\
\ts{12pt}Starting with the Lagrangian $L$, we have constructed the noncommutative quantum system $\mcS^*=(\mcC^*,H^*)$ constrained to any curved space $M^{N-1}$ in $R^N$ {\it strictly}, through the sequential projections of the system with the {\it ACCS}-expansion formalism.\\
\ts{12pt}Then, we have shown that the resultant system $\mcS^*$ is defined with the two kinds of the constrained quantum systems, $\mcS^*_{\mbtn{I}}$ and $\mcS^*_{\mbtn{II}}$, which are equivalent with each other through the projection conditions (4.35b), (4.48b) and (4.52b).\\
\ts{12pt}There, we have proved that the resultant Hamiltonians $H^*_{\mbtn{I}}$ in $\mcS^*_{\mbtn{I}}$ and $H^*_{\mbtn{II}}$ in $\mcS^*_{\mbtn{II}}$ contain the quantum correction terms in the form of the power-series of $\hbar^{2n}$ $(n\geq 1)$, which are completely missed in the usual approach with the Dirac-bracket quantization\cite{S9,S10}. \\
\ts{12pt}We have thus constructed the noncommutative quantum systems on a curved space in the {\it exact} form. 
\vs{12pt}\\

\appendix

\ts{-24pt}{\bf \LARGE Appendix}

\section{Explicit representation of $\Qhat^{\mbtn{(1)}}H$}

In the formula

$$
\begin{array}{l}

\Qhat^{\mbtn{(1)}}H\vs{6pt}\\

=-\dps{\f{\hbar}4}N+\mcU_{\mbtn{I}}(x)+\symmp{\mcU^{kl}_{\mbtn{II}}(x)}{v_kv_l}

+\symmp{\mcU_{\mbtn{III}}(x)}{p_{\lambda}}+\symmp{\symmp{\mcU^{kl}_{\mbtn{IV}}(x)}{v_kv_l}}{p_{\lambda}},
\end{array}
$$
the explicit forms of $\mcU_{\mbtn{I}}(x),\mcU_{\mbtn{II}}^{kl}(x),\mcU_{\mbtn{III}}(x)$ and $\mcU_{\mbtn{IV}}^{kl}(x)$ are given as follows:

$$
\begin{array}{lcl}

\mcU_{\mbtn{I}}(x)&=&\dps{\f{\hbar^2}4\sum^{\infty}_{n=0}\f1{n!}\left(\f{\hbar}4\right)^n(\mu^{\mbtn{(2)}}_{ikk}(x)_{;i(aa)^n}+\f{\hbar}2\mu^{\mbtn{(2)}}_{ikl}(x)_{;ikl(aa)^n})},\vs{12pt}\\

&-&\dps{\f{\hbar^2}4\sum^{\infty}_{n=1}\f1{n!}\left(\f{\hbar}4\right)^n\mu^{\mbtn{(3)}}_{kl}(x)_{;(aa)^n}G_{kl}(x)}+\dps{\f{\hbar}2\sum^{\infty}_{n=1}\f1{n!}\left(\f{\hbar}4\right)^n(\mcA^{\mbtn{(n)}}_{kk}(x)+\f{\hbar}2\mcA^{\mbtn{(n)}}_{kl}(x)_{;kl})},\vs{12pt}\\

\mcU^{kl}_{\mbtn{II}}(x)&=&\dps{\f{\hbar^2}4\sum^{\infty}_{n=0}\f1{n!}\left(\f{\hbar}4\right)^n\mu^{\mbtn{(2)}}_{ikl}(x)_{;i(aa)^n}+\sum^{\infty}_{n=1}\f1{n!}\left(\f{\hbar}4\right)^n\mcA^{\mbtn{(n)}}_{kl}(x)},\vs{12pt}\\

\mcU_{\mbtn{III}}(x)&=&\dps{\f{\hbar}2\sum^{\infty}_{n=0}\f1{n!}\left(\f{\hbar}4\right)^n\mu^{\mbtn{(3)}}_{kk}(x)_{;(aa)^n}+\f{\hbar^2}4\sum^{\infty}_{n=1}\f1{n!}\left(\f{\hbar}4\right)^n\mu^{\mbtn{(3)}}_{kl}(x)_{;kl(aa)^n}}\vs{12pt}\\

\mcU^{kl}_{\mbtn{IV}}(x)&=&\dps{\sum^{\infty}_{n=1}\f1{n!}\left(\f{\hbar}4\right)^n\mu^{\mbtn{(3)}}_{kl}(x)_{;(aa)^n}},

\end{array}
\eqno{(\mbox{A1a})}
$$
where
$$
\begin{array}{l}

\dps{\mcA^{\mbtn{(n)}}_{kl}(x)=\sum^{n}_{m=1}\mcA^{\mbtn{(n)kl}}_{\mbtn{m}}(x)},\vs{12pt}\\

\mcA^{\mbtn{(n)kl}}_{\mbtn{m}}(x)=-(2\mu^{\mbtn{(3)}}_{kl}(x)_{;(aa)^{m-1}a_m}G(x)_{;a_m}+\mu^{\mbtn{(3)}}_{kl}(x)_{;(aa)^{m-1}}G(x)_{a_ma_m})_{;(a_{m+1}a_{m+1}\cdots a_na_n)}.

\end{array}
\eqno{(\mbox{A1b})}
$$

\section{Explicit representation of $\Qhat^{\mbtn{(2)}}H^{{\mbtn{(1)}}}$}

$\Qhat^{\mbtn{(2)}}H^{\mbtn{(1)}}$ is 
$$
\begin{array}{lcl}

\Qhat^{\mbtn{(2)}}H^{\mbtn{(1)}}&=&\Qhat^{\mbtn{(2)}}\dps{\f12}\symmp{v_i}{v_i}+\Qhat^{\mbtn{(2)}}U_{\mbtn{I}}(x)+\Qhat^{\mbtn{(2)}}\symmp{U_{\mbtn{II}}^{ij}(x)}{v_iv_j}\vs{12pt}\\

&+&\Qhat^{\mbtn{(2)}}\symmp{U_{\mbtn{III}}(x)}{p_{\lambda}}+\Qhat^{\mbtn{(2)}}\symmp{\symmp{U_{\mbtn{IV}}^{ij}(x)}{v_iv_j}}{p_{\lambda}}.
\end{array}
\eqno{(\mbox{B1})}
$$

\subsection{$\Qhat^{\mbtn{(2)}}\dps{\f12}\symmp{v_i}{v_i}$}

\ts{12pt}From $(\pimm{(2)})^mv_i(x)=0$ $(m\geq 2)$ in the ACCS expansion, $\Qhat^{\mbtn{(2)}}\symmp{v_i}{v_i}$ is obtained in the following way:
$$
\begin{array}{l}

\Qhat^{\mbtn{(2)}}\dps{\f12}\symmp{v_i}{v_i}\vs{24pt}\\

=\dps{\f12\sum^{\infty}_{n= 0}\f1{n!}\left(\f{\hbar}4\right)^{n+1}\Phat^{\mbtn{(2)}}(\ximm{(2)})^{2n}(\pimm{(2)})^2\symmp{v_i}{v_i}}+\dps{\f12\sum^{\infty}_{n= 1}\f1{n!}\left(\f{\hbar}4\right)^n\Phat^{\mbtn{(2)}}(\ximm{(2)})^{2n}\symmp{v_i}{v_i}}\vs{24pt}\\

=\dps{\f{\hbar}4\sum^{\infty}_{n= 0}\f1{n!}\left(\f{\hbar}4\right)^n\Phat^{\mbtn{(2)}}(\ximm{(2)})^{2n}\mcG(x)}\vs{12pt}\\

+\dps{\f12\sum^{\infty}_{n= 1}\f1{n!}\left(\f{\hbar}4\right)^n\Phat^{\mbtn{(2)}}\sum^{2n}_{m=0}   {}_{2n}C_m\symmp{(\ximm{(2)})^mv_i}{(\ximm{(2)})^{2n-m}v_i}}\vs{24pt}\\

=\dps{\f12}\symmp{B^{ij}(x)}{\symmp{v_i}{v_j}}
+\dps{\f{\hbar}4\sum^{\infty}_{n= 0}\f1{n!}\left(\f{\hbar}4\right)^n(\nu(x)\partial)^{2n}\mcG(x)}+\dps{\f12}B^{ij}(x)U^v_{ij}(x)\vs{12pt}\\

-\dps{\f{\hbar^2}8((\nu(x)\partial)^2B^{ij}(x))G_i(x)G_j(x)+\f{\hbar^2}4G_i(x)P_{jk}(x)((\nu(x)\partial)B^{ij}(x))_{;k}}
+\dps{\f{\hbar^2}8\mcB_{ii}(x)}

\end{array}
\eqno{(\mbox{B2a})}
$$
where
$$
\begin{array}{l}

\dps{B^{ij}(x)=\sum^{\infty}_{n= 1}\f1{n!}\left(\f{\hbar}4\right)^nB^{(2n)ij}_{kk}(x)},\hs{24pt}
\dps{B^{(2n)kl}_{ij}(x)=\sum^{2n}_{m=0} {}_{2n}C_m\vLd^{(m)}_{ik}(x)
\vLd^{(2n-m)}_{jl}(x)},\vs{12pt}\\

\dps{\mcB_{ij}(x)=\sum^{\infty}_{n= 1}\f1{n!}\left(\f{\hbar}4\right)^n\mcB^{(2n)}_{ij}(x)},\vs{12pt}\\

\dps{\mcB^{(2n)}_{ij}(x)=\sum^{2n}_{m=0}  {}_{2n}C_m\left((\vLd^{(m)}_{ik}(x)
\vLd^{(2n-m)}_{jl}(x))_{;kl}+\vLd^{(m)}_{ik}(x)_{;l}
\vLd^{(2n-m)}_{jl}(x)_{;k}\right)}

\end{array}
\eqno{(\mbox{B2b})}
$$
with
$$
\begin{array}{l}

\vLd^{(n)}_{ij}(x)=(\nu(x)\partial)\vLd^{(n-1)}_{ij}(x)+\vLd^{(n-1)}_{ik}(x)\mcV_{kj}(x)\hs{12pt}(n=1,2,3,\cdots),\vs{12pt}\\

\vLd^{(0)}_{ij}(x)=\delta_{ij},\hs{12pt}\mcV_{ij}(x)=-2\mu^{{\mbtn{(2)}}}_{j;ik}(x)\nu_k(x)-\mu^{{\mbtn{(3)}}}_{ij}(x).

\end{array}
\eqno{(\mbox{B2c})}
$$

\subsection{$\Qhat^{\mbtn{(2)}}U_{\mbtn{I}}(x)$}

\ts{12pt}From $\ximm{(2)}x^i=0$, $\Qhat^{\mbtn{(2)}}U_{\mbtn{I}}(x)$ becomes
$$
\Qhat^{\mbtn{(2)}}U_{\mbtn{I}}(x)=\sum^{\infty}_{n= 1}\f1{n!}\left(\f{\hbar}4\right)^n\Phat^{\mbtn{(2)}}(\ximm{(2)})^{2n}U_{\mbtn{I}}(x)=\sum^{\infty}_{n= 1}\f1{n!}\left(\f{\hbar}4\right)^n(\nu(x)\partial)^{2n}U_{\mbtn{I}}(x).
\eqno{(\mbox{B3})}
$$

\subsection{$\Qhat^{\mbtn{(2)}}\symmp{U_{\mbtn{II}}^{ij}(x)}{v_iv_j}$}

As well as in $\Qhat^{\mbtn{(2)}}\dps{\f12}\symmp{v_i}{v_i}$, $\Qhat^{\mbtn{(2)}}\symmp{U_{\mbtn{II}}^{ij}(x)}{v_iv_j}$ is obtained as follows:
$$
\begin{array}{l}

\Qhat^{\mbtn{(2)}}\symmp{U_{\mbtn{II}}^{ij}(x)}{v_iv_j}=\Qhat^{\mbtn{(2)}}\symmp{U_{\mbtn{II}}^{ij}(x)}{\symmp{v_i}{v_j}}\vs{24pt}\\

=\dps{\sum^{\infty}_{n= 0}\f1{n!}\left(\f{\hbar}4\right)^{n+1}\Phat^{\mbtn{(2)}}(\ximm{(2)})^{2n}(\pimm{(2)})^2\symmp{U_{\mbtn{II}}^{ij}(x)}{\symmp{v_i}{v_j}}}\vs{12pt}\\

+\dps{\sum^{\infty}_{n=1}\f1{n!}\left(\f{\hbar}4\right)^n\Phat^{\mbtn{(2)}}(\ximm{(2)})^{2n}\symmp{U_{\mbtn{II}}^{ij}(x)}{\symmp{v_i}{v_j}}}\vs{24pt}\\

=\symmp{\mcM_{\mbtn{II}}^{ij}(x)}{\symmp{v_i}{v_j}}\vs{12pt}\\

+\dps{\f{\hbar}2\sum^{\infty}_{n= 0}\f1{n!}\left(\f{\hbar}4\right)^n(\nu(x)\partial)^{2n}(U_{\mbtn{II}}^{ij}(x)G_i(x)G_j(x))}+\mcM_{\mbtn{II}}^{ij}(x)U^v_{ij}(x)\vs{12pt}\\

-\dps{\f{\hbar^2}4G_i(x)G_j(x)(\nu(x)\partial)^2\mcM_{\mbtn{II}}^{ij}(x)+\f{\hbar^2}2G_i(x)P_{jk}(x)((\nu(x)\partial)\mcM_{\mbtn{II}}^{ij}(x))_{;k}}\vs{12pt}\\

+\mcU^{\mbtn{(II)}}(x),

\end{array}
\eqno{(\mbox{B4a})}
$$
where
$$
\mcM_{\mbtn{II}}^{ij}(x)=\dps{\sum^{\infty}_{n= 1}\f1{n!}\left(\f{\hbar}4\right)^n\sum^{2n}_{m=0} {}_{2n}C_m((\nu(x)\partial)^{2n-m}U_{\mbtn{II}}^{kl}(x))B^{(m)ij}_{kl}(x)},
\eqno{(\mbox{B4b})}
$$
and
$$
\begin{array}{l}

\mcU^{\mbtn{(II)}}(x)=\dps{\f{\hbar^2}2\sum^{\infty}_{n= 1}\f1{n!}\left(\f{\hbar}4\right)^n\sum^{2n}_{m=0}   {}_{2n}C_m((\nu(x)\partial)^{2n-m}U_{\mbtn{II}}^{kl}(x))_{;i}B^{(m)ij}_{kl}(x)_{;j}}\vs{12pt}\\

+\dps{\f{\hbar^2}4\sum^{\infty}_{n= 1}\f1{n!}\left(\f{\hbar}4\right)^n\sum^{2n}_{m=0}   {}_{2n}C_m((\nu(x)\partial)^{2n-m}U_{\mbtn{II}}^{ij}(x))\mcB_{ij}^{(m)}(x)}.

\end{array}
\eqno{(\mbox{B4c})}
$$

\subsection{$\Qhat^{\mbtn{(2)}}\symmp{U_{\mbtn{III}}(x)}{p_{\lambda}}$}

\ts{12pt}From $p_{\lambda}=\pi^{\mbtn{(2)}}$ and $\pimm{(2)}x^i=0$,  $\Qhat^{\mbtn{(2)}}\symmp{U_{\mbtn{III}}(x)}{p_{\lambda}}$ is given as follows:
$$
\begin{array}{l}

\Qhat^{\mbtn{(2)}}\symmp{U_{\mbtn{III}}(x)}{p_{\lambda}}=\Qhat^{\mbtn{(2)}}\pipp{(2)}U_{\mbtn{III}}(x)
=\dps{\sum^{\infty}_{n= 1}\f1{n!}\left(\f{\hbar}4\right)^n\Phat^{\mbtn{(2)}}(\ximm{(2)})^{2n}\pipp{(2)}U_{\mbtn{III}}(x)}\vs{12pt}\\

=\dps{\f{\hbar}2\sum^{\infty}_{n=0}\f1{n!}\left(\f{\hbar}4\right)^n(\nu(x)\partial)^{2n+1}U_{\mbtn{III}}(x)=\f{\hbar}2\sum^{\infty}_{n=0}\f1{n!}\left(\f{\hbar}4\right)^n(\nu(x)\partial)^{2n}(\nu_k(x)U_{\mbtn{III}}(x)_{;k}).}

\end{array}
\eqno{(\mbox{B5})}
$$

\subsection{$\Qhat^{\mbtn{(2)}}\symmp{\symmp{U_{\mbtn{IV}}^{ij}(x)}{v_iv_j}}{p_{\lambda}}$}

\ts{12pt}As well as in $\Qhat^{\mbtn{(2)}}\symmp{U_{\mbtn{II}}^{ij}(x)}{v_iv_j}$ and $\Qhat^{\mbtn{(2)}}\symmp{U_{\mbtn{III}}(x)}{p_{\lambda}}$, $\Qhat^{\mbtn{(2)}}\symmp{\symmp{U_{\mbtn{IV}}^{ij}(x)}{v_iv_j}}{p_{\lambda}}$ is obtained in the following way:\\
$$
\begin{array}{l}

\Qhat^{\mbtn{(2)}}\symmp{\symmp{U_{\mbtn{IV}}^{ij}(x)}{v_iv_j}}{p_{\lambda}}=\Qhat^{\mbtn{(2)}}\pipp{(2)}\symmp{U_{\mbtn{IV}}^{ij}(x)}{\symmp{v_i}{v_j}}\vs{24pt}\\

=\dps{\sum^{\infty}_{n= 1}\f1{n!}\left(\f{\hbar}4\right)^{n+1}\Phat^{\mbtn{(2)}}(\ximm{(2)})^{2n}(\pimm{(2)})^2\left(\pipp{(2)}\symmp{U_{\mbtn{IV}}^{ij}(x)}{\symmp{v_i}{v_j}}\right)}\vs{12pt}\\

+\dps{\sum^{\infty}_{n= 1}\f1{n!}\left(\f{\hbar}4\right)^n\Phat^{\mbtn{(2)}}(\ximm{(2)})^{2n}\left(\pipp{(2)}\symmp{U_{\mbtn{IV}}^{ij}(x)}{\symmp{v_i}{v_j}}\right)}\vs{24pt}\\

=\dps{\f{\hbar}2\symmp{\mcM_{\mbtn{IV}}^{ij}(x)}{\symmp{v_i}{v_j}}}\vs{12pt}\\

+\dps{\f{\hbar^2}4\sum^{\infty}_{n= 0}\f1{n!}\left(\f{\hbar}4\right)^n(\nu(x)\partial)^{2n+1}(U_{\mbtn{IV}}^{ij}(x)G_i(x)G_j(x))+\f{\hbar}2\mcM_{\mbtn{IV}}^{ij}(x)U^v_{ij}(x)}\vs{12pt}\\

-\dps{\f{\hbar^3}8G_i(x)G_j(x)(\nu(x)\partial)^2\mcM_{\mbtn{IV}}^{ij}(x)+\f{\hbar^3}4G_i(x)P_{jk}(x)((\nu(x)\partial)\mcM_{\mbtn{IV}}^{ij}(x))_{;k}}\vs{12pt}\

+\mcU^{\mbtn{(IV)}}(x)

\end{array}
\eqno{(\mbox{B6a})}
$$
where
$$
\mcM_{\mbtn{IV}}^{ij}(x)=\dps{\sum^{\infty}_{n= 0}\f1{n!}\left(\f{\hbar}4\right)^n\sum^{2n+1}_{m=0} {}_{2n+1}C_m((\nu(x)\partial)^{2n+1-m}U_{\mbtn{IV}}^{kl}(x))B^{(m)ij}_{kl}(x)}.
\eqno{(\mbox{B6b})}
$$
and
$$
\begin{array}{l}

\mcU^{\mbtn{(IV)}}(x)=\dps{\f{\hbar^3}4\sum^{\infty}_{n= 0}\f1{n!}\left(\f{\hbar}4\right)^n\sum^{2n+1}_{m=0}   {}_{2n+1}C_m((\nu(x)\partial)^{2n+1-m}U_{\mbtn{IV}}^{ij}(x))_{;k}B^{(m)kl}_{ij}(x)_{;l}}\vs{12pt}\\

+\dps{\f{\hbar^3}8\sum^{\infty}_{n= 0}\f1{n!}\left(\f{\hbar}4\right)^n\sum^{2n+1}_{m=0}   {}_{2n+1}C_m((\nu(x)\partial)^{2n-+1m}U_{\mbtn{IV}}^{ij}(x))\mcB_{ij}^{(m)}(x)}.

\end{array}
\eqno{(\mbox{B6c})}
$$


\newpage

\begin{thebibliography}{99}

\bibitem{S1} For example,\\
E. M. C. Abreu, R. Amorim and W. Guzm$\acute{\mbox{a}}$n Ram$\acute{\mbox{i}}$rez, JHEP {\bf 1103} (2011) 135;\\
{\it  Noncommutative Particles in Curved Spaces}, arXiv:1011.0023v2 [hep-th] 17 Nov 2010;\\
M. Buri$\acute{\mbox{c}}$, M. Dimitrijevi$\acute{\mbox{c}}$, V. Radovanovi$\acute{\mbox{c}}$ and M. Wohlgenannt, {\it Quantization of a gauge theory on a curved noncommutative space},  arXiv:1203.3016v1 [hep-th] 14 Mar 2012;\\
V. G. Kupriyanov, {\it Hydrogen atom on curved noncommutative space}, arXiv:1209.6105v2 [math-ph] 5 Jun 2013;\\
J. Barcelos-Neto, {\it  Noncommutative fields in curved space}, arXiv:hep-th/0212094v1 9 Dec 2002.

\bibitem{S2} M. Nakamura, {\it Alternative Approach to Noncommutative Quantum Mechanics on a Curved Space}, arXiv:1512.00143v5 [hep-th] 5 Oct 2016; and References there in.

\bibitem{S3} M. Nakamura, {\it Star-product Description of Quantization in Second-class Constraint Systems}, arXiv:1108.4108v6 [math-ph] 13 Sep 2014.

 \bibitem{S4} M. Nakamura and N. Mishima, Prog.Theor.Phys. {\bf 81} (1989) 514;\\
M. Nakamura and H. Minowa, J.Math.Phys. {\bf 34} (1993) 50.

\bibitem{S5} I. A. Batalin and E. S. Fradkin, Nucl.Phys. {\bf B279} (1987) 145.

\bibitem{S6} I. A. Batalin, S. L. Lyakhovich and R. Marnelius, {\it Projection operator approach to general constrained systems}, arXiv:hep-th/0112175v2 12 Mar 2002. 

\bibitem{S7} M. Nakamura, {\it Uncertainty Relations and Quantum Effects of Constraints in Chern-Simons Theory}, arXiv:1302.4520v2 [hep-th] 22 Feb 2013.

\bibitem{S8} M. Nakamura and K. Kojima, {\it Coherent states in constrained systems}, Nuovo Cim. {\bf 116B}, N.3 (2001) 287.

 \bibitem{S9} P. A. M. Dirac, {\it Lectures on Quantum Mechanics} (Belfer Graduate School of Science, Yeshiba University, New York) 1969.
 
\bibitem{S10} A. G. Hanson, T. Regge and C. Teitelboim, {\it Constrained Hamiltonian Systems} (Accademia Nazionale dei Lincei Roma) 1976.

\end{thebibliography}
\end{document}